\documentclass[structabstract]{aa}  
\usepackage{graphicx}
\usepackage{txfonts}

\usepackage{upgreek}
\usepackage{color}
\usepackage{url}
\usepackage{multirow}
\usepackage{subfig}
\usepackage{lscape}
\usepackage{rotating}

\usepackage{natbib}
\bibpunct{(}{)}{;}{a}{}{,} % to follow the A&A style

\makeatletter
\DeclareRobustCommand*\textsubscript[1]{%
    \@textsubscript{\selectfont#1}}
  \newcommand{\@textsubscript}[1]{%
    {\m@th\ensuremath{_{\mbox{\fontsize\sf@size\z@#1}}}}}
\makeatother

\newcommand{\irac}{IRAC}

\newcommand{\pacs}{PACS}
\newcommand{\spire}{SPIRE}
\newcommand{\hst}{HST}
\newcommand{\hawki}{\mbox{HAWK-I}}

\newcommand{\laboca}{LABOCA}
\newcommand{\herschel}{\textit{Herschel}}
\newcommand{\spitzer}{\textit{Spitzer}}
\newcommand{\chandra}{\textit{Chandra}}
\newcommand{\ego}{EGO}
\newcommand{\cgo}{CGO}
\renewcommand{\mho}{MHO}
\newcommand{\sed}{SED}
\newcommand{\psf}{PSF}
\newcommand{\fwhm}{FWHM}
\newcommand{\rgb}{RGB}
\newcommand{\yso}{YSO}
\newcommand{\apex}{APEX}
\newcommand{\mopex}{MOPEX}
\newcommand{\cnc}{CNC}
\newcommand{\hh}{HH}
\newcommand{\hhc}{HHc-}

\newcommand{\ie}{i.\,e.}
\newcommand{\eg}{e.\,g.}

\newcommand{\timestento}[1]{\ensuremath{\cdot 10^{#1}}}
\newcommand{\mum}{\,\ensuremath{\upmu\mathrm{m}}}

\begin{document}
   \title{Jet-driving protostars identified from infrared observations\thanks{
          This work is based in part on observations made with the \spitzer\ Space Telescope, which is operated by the Jet Propulsion Laboratory, California Institute of Technology, under a contract with NASA, and on data collected by \herschel, an ESA space observatory with science instruments provided by European-led Principal Investigator consortia and with important participation from NASA.
         } of the Carina Nebula complex}

   \author{H. Ohlendorf
          \inst{1}
           \and
           T. Preibisch
           \inst{1}
           \and
           B. Gaczkowski
           \inst{1}
           \and
           T. Ratzka
           \inst{1}
           \and
           R. Grellmann
           \inst{1}
           \and
           A. F. McLeod
           \inst{1}
          }

   \institute{Universit\"ats-Sternwarte M\"unchen, Ludwig-Maximilians-Universit\"at,
             Scheinerstra{\ss}e 1, 81679 M\"unchen, Germany \\
              \email{ohlendorf@usm.uni-muenchen.de}
             }

   \date{Received 30 September, 2011; accepted 19 January, 2012}

\abstract
  % context heading (optional)
   {}
  % aims heading (mandatory)
   {Jets are excellent signposts for very young embedded protostars, so we want to identify jet-driving protostars as a tracer of the currently forming generation of stars in the Carina Nebula, which is one of the most massive galactic star-forming regions and which is characterised by particularly high levels of massive-star feedback on the surrounding clouds.}
  % methods heading (mandatory)
   {We used archive data to construct large $(\ga 2\degr \times 2\degr$) \spitzer\ \irac\ mosaics of the Carina Nebula and performed a spatially complete search for objects with excesses in the 4.5\,\mum\ band, typical of shock-excited molecular hydrogen emission. We also identified the mid-infrared point sources that are the likely drivers of previously discovered Herbig-Haro jets and molecular hydrogen emission line objects. We combined the \spitzer\ photometry with our recent \herschel\ far-infrared data to construct the spectral energy distributions, and used the Robitaille radiative-transfer modelling tool to infer the properties of the objects.}
  % results heading (mandatory)
   {The radiative-transfer modelling suggests that the jet sources are protostars with masses between $\sim 1\,M_\odot$ and $\sim 10\,M_\odot$ that are surrounded by circumstellar disks and embedded in circumstellar envelopes.}
  % conclusions heading (optional)
   {The estimated protostar masses $\leq 10\,M_\odot$ suggest that the current star-formation activity in the Carina Nebula is restricted to low- and intermediate-mass stars. More optical than infrared jets can be observed, indicating that star formation predominantly takes place close to the surfaces of clouds.}


  \maketitle

%________________________________________________________________

\section{Introduction}
\label{sec:intro}

Most stars in the Galaxy are born in massive star-forming regions. The high-mass stars profoundly influence their environments by their strong ionising radiation and powerful stellar winds that can disperse the natal molecular clouds.
However, ionisation fronts and expanding wind-driven superbubbles can also compress surrounding clouds, thereby possibly triggering the formation of new generations of stars. Although this feedback is fundamental for understanding the star-formation processes, the details of the astrophysical processes at work are  still not understood well, mainly because regions with high levels of massive-star feedback are usually too far away for detailed studies.

The Great Nebula in Carina \citep[NGC 3372; see][for an overview and full references]{smith2008} provides a unique target for studies of massive-star feedback. The Carina Nebula complex (hereafter \cnc) is located at a moderate and very well-determined distance of 2.3\,kpc. With 70 known O-type and Wolf-Rayet stars \citep{smith2006} -- among them several of the most massive  ($M \ga 100\,M_\odot$) and luminous stars known in our Galaxy -- it represents the nearest southern region with a large massive stellar population.

The presence of these very massive stars implies a very high level of stellar feedback in the \cnc. At the same time, the comparatively small distance to the \cnc\ guarantees that we still can study details of the cluster and cloud structure at sufficient spatial resolution so as to detect and characterise the low-mass stellar populations.

During the past few years, several large surveys of the \cnc\ have been performed. The \chandra\ X-ray observatory obtained a deep wide-field ($\sim 1.4\,\deg^2$) X-ray survey of the Carina complex \citep[see][]{townsley2011} that led to the detection of 14\,368 X-ray point sources \citep{broos2011_catalogue}, as well as copious amounts of diffuse X-ray emission \citep{townsley2011_diffuse}. The analysis of the X-ray and infrared properties of the point sources suggested that 10\,714 objects are young stars in the \cnc\ \citep{broos2011_bayes}, providing, for the first time, a large (but luminosity-limited) sample of the young stellar population in the area.

A very deep near-infrared survey of the central 1280 square-arcminute area performed with \hawki\ at the ESO 8\,m Very Large Telescope \citep{preibisch2011_hawki} has revealed more than 600\,000 infrared sources. The combination of this infrared catalogue with the list of X-ray sources provides new insight into the properties of the stellar populations in the \cnc\ \citep{preibisch2011_cccp}.

\spitzer\ infrared surveys located numerous embedded young stellar objects (\yso s) throughout the \cnc\ \citep{smith2010_spitzer, povich2011}.
A large ($\sim 1.5\,\deg^2$) sub-mm survey of the \cnc\ \citep{preibisch2011_laboca} provided essential information about the structure and properties of the cold clouds.

In combination with previous results, these new surveys have yielded very interesting information about the star-formation process in the Carina Nebula. The feedback of the numerous massive stars has already largely dispersed the original
molecular clouds in the central region of the Carina Nebula. A few parsecs from the centre, however, large amounts of rather dense clouds are still present. Infrared images show how these clouds are currently eroded and shaped by the radiation and winds from
the massive stars, giving rise to numerous giant dust pillars, especially in the so-called ``South Pillars'' region \citep{smith2010_spitzer}, but also in the clouds northwest of $\upeta$\,Car. 
Several very young stellar objects \citep[\eg][]{mottram2007} and a spectacular young embedded cluster \citep[the ``Treasure Chest Cluster''; see][]{smith2005_treasurechest} have been found in the molecular clouds, providing clear evidence that
vigorous star formation activity is going on in these irradiated clouds.
It is thought that the formation of these objects was triggered by the advancing ionisation  fronts that originate from the (several Myr old) high-mass stars in the \cnc\ \citep{smith2010_spitzer, povich2011}. While this scenario appears very reasonable and is supported by recent numerical simulations of the evolution of irradiated clouds \citep[\eg][]{gritschneder2010}, the existing data cannot provides clear proof of the triggered--star-formation scenario.
The alternative explanation that stars form spontaneously everywhere in the clouds (\ie\ independent of external influences) and are only revealed by the advancing ionisation fronts (that simply remove the clouds surrounding the newly formed stars) cannot be
ruled out. If it was possible to show that star formation is spatially restricted to the irradiated surface of the
clouds (where the radiation pressure is highest), this would provide a very substantial argument in favour of the triggered star formation scenario.

\begin{figure*}
\centering
\includegraphics[width=\textwidth]{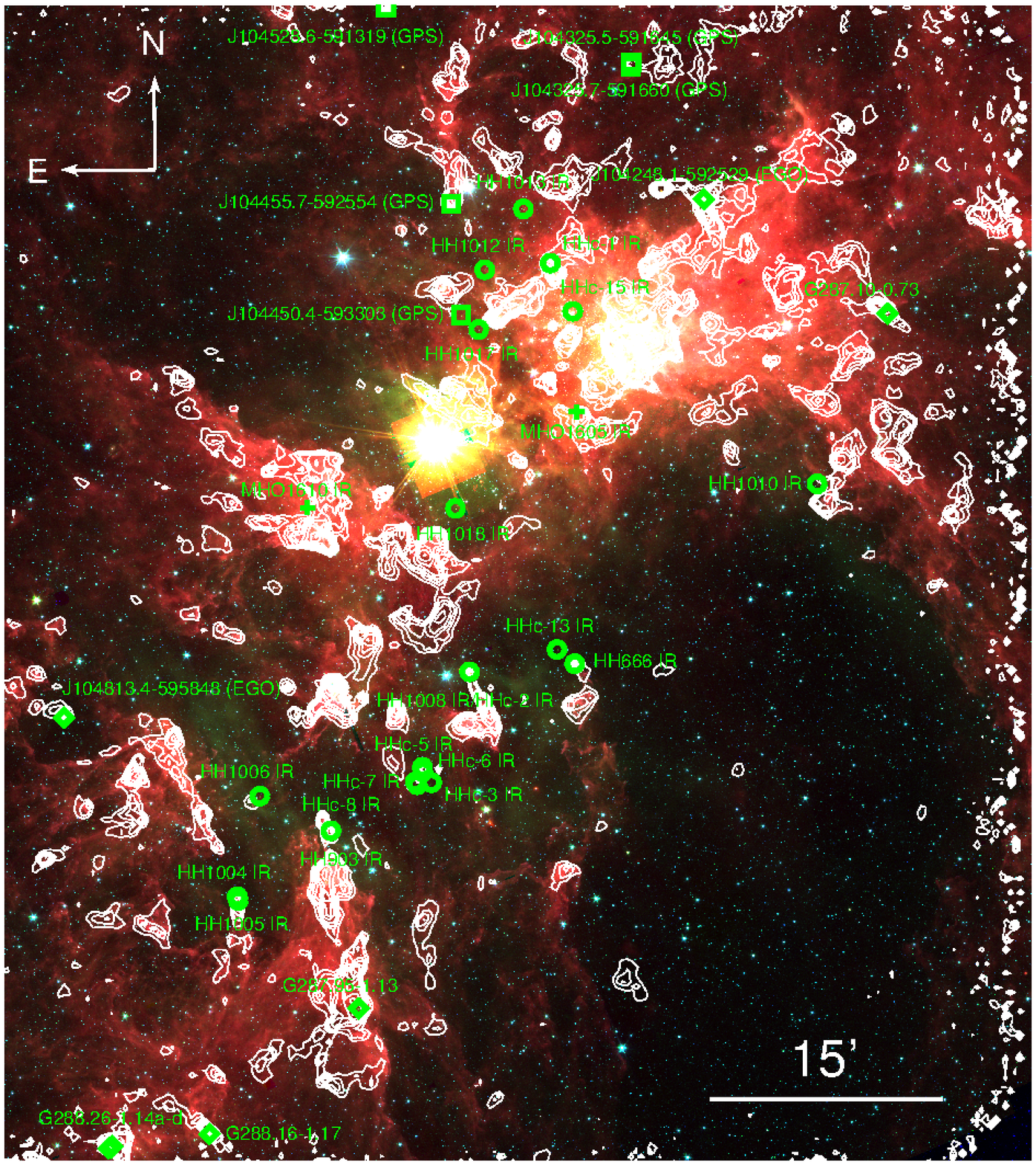}
  \caption{\spitzer\ \rgb\ image with overlaid \laboca\ 870\mum\ contours, shown for the same field as in \citet{smith2010_jets} for identifying Herbig-Haro jets. The first three contour levels are 0.06, 0.12, and 0.18 Jy/beam, while the further levels increase by a factor of the square root of two. The jet sources in the region are marked in green. Circles indicate sources of \hh\ jets and candidates, diamonds those of \ego s, and squares compact green objects. The sources of \mho s are marked by crosses.}
  \label{fig:overview}
\end{figure*}

However, the results so far do not allow drawing strong conclusions about the spatial distribution of the currently forming stars.
While the infrared data are deep enough to detect the young stellar objects in the CNC easily, major problems arise from the very high level of field-star contamination that is a consequence of the \cnc's location almost exactly on the galactic plane.
Only a few percent of all infrared sources seen in the \spitzer\ and \hawki\ images are actually young stellar objects related to the \cnc, while the vast majority are background contaminants \citep[see discussion in][]{smith2010_spitzer, povich2011, preibisch2011_hawki}.

The main problem in any study of the star-formation process in the \cnc\ thus is not to detect the \yso s, but to discern \yso s in the \cnc\ from background sources. In this context, protostellar jets provide a very good signpost for \yso s, as they point towards objects in the latest generation of the currently forming stars \citep[see][]{bally2007}. The spatial distribution of jet-driving protostars can thus provide very important information about the regions where star formation presently occurs. These protostars are so young that they are still located very close to their birth site, which means that the jet-driving \yso s trace the spatial location of the currently forming \yso\ population. 

In the strong shocks within the jet flows, or at locations where the jet impacts surrounding clouds, the gas is strongly heated and collisionally excited. This can result in strong, shock-excited line emission, depending on the state of the flowing material and the surrounding clouds. Jets propagating in the diffuse, \textit{atomic} gas outside the dense molecular clouds often produce prominent \textit{atomic} emission lines, such as H$\alpha$ or [SII], that can be easily observed in optical images and are known as Herbig-Haro (HH) objects \citep[see][]{reipurth2001}. 
If, on the other hand, the jet propagates within a dense molecular cloud, emission from collisionally excited molecular hydrogen is often seen. The radiative decay of the excited molecules via ro-vibrational transitions produces strong emission lines in the near- and mid-infrared wavelength range.

The 2.12\mum\  $\nu=1\!-\!0$\, S(1) ro-vibrational emission line of molecular hydrogen is one widely used tracer of these
shocks \citep[\eg,][]{mccaughrean1994, smith2007, davis2008}. The \spitzer\ \irac\ band 2, centred at 4.5\,\mum, contains several strong H$_2$ lines, including the transitions $\nu=2\!-\!1$\,O(7), $\nu=0\!-\!0$\,S(11), $\nu=0\!-\!0$\,S(10), $\nu=2\!-\!1$\,O(8), 
$\nu=1\!-\!0$\,O(9), $\nu=0\!-\!0$\,S(9), and $\nu=0\!-\!0$\,S(8). This is why jets often appear as prominent green nebulae in \spitzer\ \rgb\ images when the 4.5\mum\ image is used for the green colour channel \citep[see \eg][]{gutermuth2008}. These objects are often denoted as ``extended green objects'' (\ego s) or ``green fuzzies''.

Three recent searches for jets have been performed in the \cnc. \citet{smith2010_jets} used the ACS camera on the Hubble Space Telescope (\hst) to perform a deep H$\upalpha$ narrow-band imaging survey of a $13.6\arcmin \times 27\arcmin$ region in the centre of the Carina Nebula, as well as several non-contiguous pointings at various positions in the Nebula. They detected 40 \hh\ jets and \hh\ jet candidates within the $\sim 0.8\degr \times 1.1\degr$ area that was partly covered by their observations (see their Fig.~1). 
\citet{smith2010_spitzer} analysed the \spitzer\ data of the South Pillars within the \cnc\ and identified four \ego s. 

In the context of the \hawki\ near-infrared survey of the central $\sim 1280$~square-arcminute region of the \cnc\ described in \citet{preibisch2011_hawki}, we used images obtained through a narrow-band filter centred on the 2.12\,\mum\  $\nu=1\!-\!0$\, S(1) ro-vibrational emission line of molecular hydrogen to search for molecular hydrogen emission-line objects (MHOs).
This resulted in the detection of six MHOs, which are listed as MHO\,1605 to 1610 in the ``Catalogue of Molecular Hydrogen Emission-Line Objects in Outflows from Young Stars''.\footnote{\url{http://www.jach.hawaii.edu/UKIRT/MHCat/}}

In this paper, we are going to extend the search for jets and their driving protostars to the full area of the \cnc\ ($\sim 1\,\deg^2$ around $\upeta$\,Car as shown in Fig.~\ref{fig:overview}).
In the first step, we use the large \spitzer\ \irac\ mosaics we have created from available archive data, which provide a spatially complete coverage of the entire extent of the \cnc, to search for more objects with 4.5\mum\ excess emission, beyond the \ego s already detected by \citet{smith2010_spitzer}.
In the second step, we identify the driving sources of the newly detected jets and the known \hh\ jets and \mho s in the \spitzer\ images (Sect.~\ref{sec:revealing}). Using the tools developed by \citet{robitaille2006, robitaille2007}, we fit the spectral energy distributions (\sed s) of the jet sources by combining \spitzer\ photometry with far-infrared fluxes derived from our very recent \herschel\ mapping and our previous LABOCA sub-mm study of the \cnc\ \citep{preibisch2011_laboca}. We derive estimates of the stellar and circumstellar parameters of the jet sources by radiative-transfer modelling of the SEDs and investigate the spatial distribution of the jet-driving protostars (Sect.~\ref{sec:sources}) to infer information about the star formation mechanisms currently at work in the Carina Nebula.

\section{Observational data}
\label{sec:data}

We used the following infrared data sets firstly to search for jet sources and secondly to determine source fluxes to construct their \sed s for the radiative transfer modelling as described in Sect.~\ref{sec:seds}.

\subsection{\spitzer\ images and photometry}
\label{sec:data-spitzer}

The data used here were taken during the \spitzer\ cold mission phase, in July 2008, with the InfraRed Array Camera \citep[\irac;][]{fazio2004} ("Galactic Structure and Star Formation in Vela-Carina" programme; PI: Steven R. Majewski, Prog-ID: 40791). They were retrieved through the \spitzer\ Heritage Archive. Using the MOsaicker and Point source EXtractor package \citep[\mopex; ][]{makovoz2005} software provided by the \spitzer\ Science Center\footnote{\url{http://irsa.ipac.caltech.edu/}} (SSC), we assembled the basic calibrated data into mosaics covering the area around $\upeta$\,Car, incorporating the whole extent of the central Carina nebula.

\addtocounter{footnote}{-1}

For these \irac\ mosaics we performed photometry using the Astronomical Point source EXtractor (\apex) module of \mopex. During the process, photometry was carried out individually for each image in the stack and combined internally to provide photometry data for the entire mosaic. Point response functions (PRFs) for this purpose are available from the SSC and the appropriate set of PRFs for the time of the observations was employed.
Outliers were removed using the \textit{Box Outlier Detection} method within \mopex\ and backgrounds between the tiles were matched using the \textit{Overlap} module before the mosaics were constructed.

Three \spitzer\ astronomical observation requests (AORs) each were merged before performing aperture photometry on the constructed mosaics for all four \irac\ bands (3.6\mum, 4.5\mum, 5.8\mum, and 8.0\mum) separately. 
Finally, with the Bandmerge program provided by the \spitzer\ Science Center\footnotemark\, an overall catalogue was created that contained all sources detected with \irac\ in the area, their positions and fluxes in all four bands, retaining the information of the detection bands per source.
Our catalogue agrees well with the catalogue constructed by \citet{smith2010_spitzer} within the area where both overlap. Source positions, as well as measured fluxes, are in good concordance between the two catalogues. The root mean square (RMS) deviation in magnitudes for sources is 0.092\,mag in the 4.5\mum\ band, 0.13\,mag in the 3.6\mum\ band and 0.17\,mag in the two longest-wavelength bands.

The one-sigma sensitivity limits for \spitzer\ for a high background as in the present case given in the IRAC Instrument Handbook\footnote{\url{http://irsa.ipac.caltech.edu/data/SPITZER/docs/irac/}} are 34, 41, 180, and $156\,\upmu$J for 3.6\mum, 4.5\mum, 5.8\mum, and 8.0\mum, respectively. The exact values depend on the background, which within our field varies strongly.

For this study, we focused on the area of around one square degree as shown in Fig.~\ref{fig:overview}, the same area as analysed in \cite{smith2010_jets}. One \ego\ slightly north of this area is included in the study.

\subsection{\herschel\ far-infrared fluxes}
\label{sec:data-herschel}

For further characterisation of the jet sources, we included fluxes extracted from the data of our recent \herschel\ far-infrared survey of the \cnc.
These observations were performed in December 2010 (Open time project, PI: Thomas Preibisch, Prog-ID: OT1-tpreibis-1), using the parallel fast scan mode at $60 \arcsec \mathrm{s}^{-1}$. With simultaneous five-band imaging with \pacs\ \citep{poglitsch2010} at 70\mum\ and 160\mum\ and \spire\ \citep{griffin2010} at 250\mum, 350\mum\ and 500\mum\ two orthogonal scan maps were obtained to cover an area of $2.3 \times 2.3 \deg^2$. A full description of these observations and the subsequent data processing will be given elsewhere. 

\herschel\ fluxes were only extracted for cases where a compact, point-like \herschel\ object could be found at the position of the \spitzer\ source. No photometry was performed in cases where strong extended emission shows up in the \herschel\ maps.
Detection and photometry of  point-like sources in \herschel\ bands were carried out with CUTEX \citep{molinari2011}. 
Fluxes down to about 1\,Jy can be detected with \herschel.
Since the angular resolution of \herschel\ with a point spread function (\psf) between 5\arcsec\ and 36\arcsec, depending on band, is considerably lower than that of \spitzer\ at an \fwhm\ of the \psf\ between 1.5\arcsec\ and 2\arcsec\ \citep{fazio2004}, source detection with \herschel\ is more difficult than it is in \spitzer\ images. In many cases a single point-like source is, however, identified as a counterpart. The remaining cases fall into two different categories: (i) those where the local emission is predominantly uniform, and no compact \herschel\ counterpart to the \spitzer\ point-like source can be detected, and (ii) those for which a counterpart is identified that is a more or less pronounced luminosity increase within a larger structure. Overall, 19 of the 37 \spitzer-identified jet sources (51\%) do not have a compact \herschel\ counterpart so no photometry was performed for them.

\subsection{LABOCA sub-mm fluxes}
\label{sec:data-laboca}

We complemented the \herschel\ photometry with 870\mum\ fluxes extracted from our \laboca\ map of the CNC \citep{preibisch2011_laboca} for those cases whenever a compact source was seen at the location of the \spitzer-identified jet sources
in the \laboca\ map.
The source fluxes were extracted from $\approx 25''$ diameter apertures centred on each object.

\subsection{HAWK-I and 2MASS}

We also inspected the Two Micron All Sky Survey \citep[2MASS;][]{skrutskie2006} images and our very deep \hawki\ near-infrared images \citep[see][]{preibisch2011_hawki} and searched for near-IR counterparts of the \spitzer-identified jet sources. When a near-IR source could be clearly identified with the \spitzer\ source, we used the $J$-, $H$-, and $K_s$-band fluxes as given in our \hawki\ photometry catalogue \citep{preibisch2011_hawki} or (for sources outside the area of the \hawki\ survey) the 2MASS Point-Source Catalog. The typical completeness limits of the \hawki\ image are $\sim21$\,mag (J), $\sim20$\,mag (H), and $\sim19$\,mag (K\textsubscript{s}) \citep{preibisch2011_cccp}.

\section{Revealing jet sources in the Carina Nebula}
\label{sec:revealing}

Since \hh\ objects and \mho s were detected in other wavelengths, while \ego s and compact green objects (\cgo s) were identified in the \spitzer\ images themselves, slightly different strategies were employed in searching for infrared sources.

\begin{figure}
\centering
\includegraphics[width=0.49\textwidth]{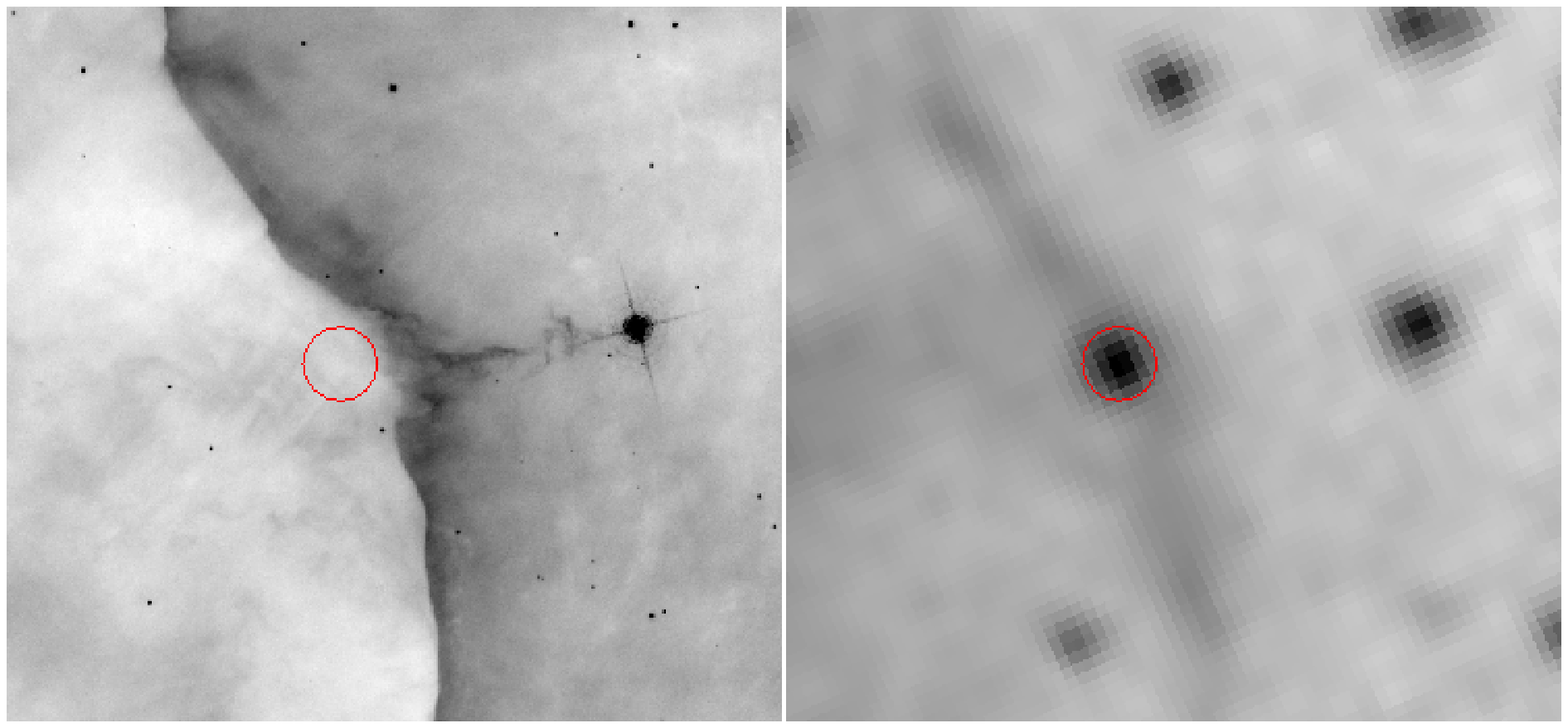}
  \caption{\hh903 is clearly visible in the \hst\ H$\upalpha$ image (left), but no source for the jet is visible. In the \spitzer\ 4.5\mum\ image (right), a point-like source (red circle) is revealed that is very likely the source of jet emission. The circle marking the source has a radius of 2\arcsec.}
  \label{fig:hh903}
\end{figure}

As an example of a case where a \spitzer\ infrared source can be identified while no source can be discerned in the optical \hst\ image, we show \hh\,903 in Fig.~\ref{fig:hh903}. Overview figures of optical, near-IR, \spitzer, and \herschel\ images can be seen in Fig.~\ref{fig:images_hhs} (\hh\ objects), Fig.~\ref{fig:images_mhos} (\mho s), Fig.~\ref{fig:images_egos} (\ego s), and Fig.~\ref{fig:images_gpqs} (\cgo s).

A list of all jet objects considered here was compiled in Table~\ref{tab:all_detections}. For all objects studied, it indicates in which wavelength bands they could be detected.

\begin{table*}
\caption[]{Infrared detections for sources of jet objects within the Carina nebula.}
\label{tab:all_detections}
\centering

\begin{tabular}{l l c c c c}
\hline\hline
\noalign{\smallskip}
Jet Object & Jet Source  & Spitzer  &   Herschel  & 2MASS & HAWKI \\
\noalign{\smallskip}
\hline
\noalign{\smallskip}
\hh666        & J104351.5--595521 & yes & yes & yes\tablefootmark{a} & --\tablefootmark{a}  \\
\hh900        &                   & no  & yes & no  & no  \\
\hh901        &                   & no  & no  & no  & yes \\
\hh902        &                   & yes & no  & no  & no  \\
\hh903        & J104556.4--600608 & yes & yes & no  & no  \\
\hh1004       & J104644.8--601021 & yes & yes & no  & no  \\
\hh1005       & J104644.2--601035 & yes & yes & no  & --  \\
\hh1006       & J104632.9--600354 & yes & no  & no  & no  \\
\hh1007       &                   & no  & no  & no  & --  \\
\hh1008       & J104445.4--595555 & yes & yes & yes & yes \\
\hh1009       &                   & no  & no  & no  & no  \\
\hh1010       & J104148.7--594338 & yes & yes & no  & --  \\
\hh1011       &                   & no  & no  & no  & no  \\
\hh1012       & J104438.7--593008 & yes & no  & no  & no  \\
\hh1013       & J104419.2--592612 & yes & no  & no  & no  \\
\hh1014       & J104545.9--594106 & yes & no  & no  & no  \\
\hh1015       &                   & no  & no  & yes & --  \\
\hh1016       &                   & no  & no  & no  & no  \\
\hh1017       & J104441.5--593357 & yes & no  & no  & no  \\
\hh1018       & J104452.9--594526 & yes & no  & no  & no  \\
\hhc1         & J104405.4--592941 & yes & no  & no  & no  \\
\hhc2         & J104445.4--595555 & yes & yes & yes & yes \\
\hhc3         & J104504.6--600303 & yes & no  & no  & no  \\
\hhc4         &                   & no  & no  & no  & no  \\
\hhc5         & J104509.4--600203 & yes & yes & no  & yes \\
\hhc6         & J104509.2--600220 & yes & no  & no  & no  \\
\hhc7         & J104513.0--600259 & yes & no  & no  & no  \\
\hhc8         & J104512.0--600310 & yes & no  & no  & no  \\ 
\hhc9         &                   & no  & no  & no  & no  \\
\hhc10        &                   & no  & no  & no  & no  \\
\hhc11        &                   & no  & no  & no  & no  \\
\hhc12        &                   & no  & no  & no  & no  \\
\hhc13        & J104400.6--595427 & yes & no  & no  & --  \\
\hhc14        &                   & no  & no  & no  & no  \\
\hhc15        & J104353.9--593245 & yes & no  & no  & no  \\ 
\ego          & J104040.5--585319 & yes & yes & no  & --  \\ 
\ego          & J104248.1--592529 & yes & yes & yes & --  \\
\ego          & J104813.4--595848 & yes & yes & no  & no  \\
G287.10-0.73  & J104114.3--593239 & yes & yes & no  & --  \\
G287.95-1.13  & J104542.0--601732 & yes & yes & no  & --  \\
G288.15-1.17  & J104700.6--602535 & no  & no  & no  & --  \\
G288.26-1.14a & J104750.3--602618 & yes & yes & no  & --  \\
G288.26-1.14b & J104752.1--602627 & yes & no  & no  & --  \\
G288.26-1.14c & J104750.9--602625 & yes & no  & no  & --  \\
G288.26-1.14d & J104751.7--602624 & yes & no  & no  & --  \\
\cgo          & J104325.5--591645 & yes & yes & no  & --  \\
\cgo          & J104325.7--591660 & yes & yes & no  & --  \\
\cgo          & J104450.4--593303 & yes & no  & no  & no  \\
\cgo          & J104455.7--592554 & yes & yes & no  & no  \\
\cgo          & J104528.6--591319 & yes & yes & yes & --  \\
MHO1605       & J104351.5--593911 & yes & no  & no & yes  \\
MHO1606       &                   & no  & no  & no  & no  \\
MHO1607       &                   & no  & no  & no  & no  \\
MHO1608       &                   & no  & no  & no  & no  \\
MHO1609       &                   & no  & no  & no  & no  \\
MHO1610       &                   & yes & yes & no  & no  \\
\hline
\end{tabular}

\tablefoot{The Herbig-Haro objects and candidates are as detected by \citet{smith2010_jets}, Molecular Hydrogen objects as identified by \citet{preibisch2011_hawki}, the Extended Green Objects bearing a number beginning with \textit{G} as detected by \citet{smith2010_spitzer}. Images of them can be found in the respective source publications. The remaining four \ego s and the compact green objects were detected from our Spitzer data. For \hawki, where neither \emph{yes} nor \emph{no} are indicated the source is located outside the surveyed field. \\ \tablefoottext{a}{The SOFI JHK photometry for HH666 IR was taken from \citet{smith2004_axisofevil}.}}

\end{table*}

\begin{figure*}
 \centering
 \begin{minipage}{\textwidth}
  \includegraphics[width=\textwidth]{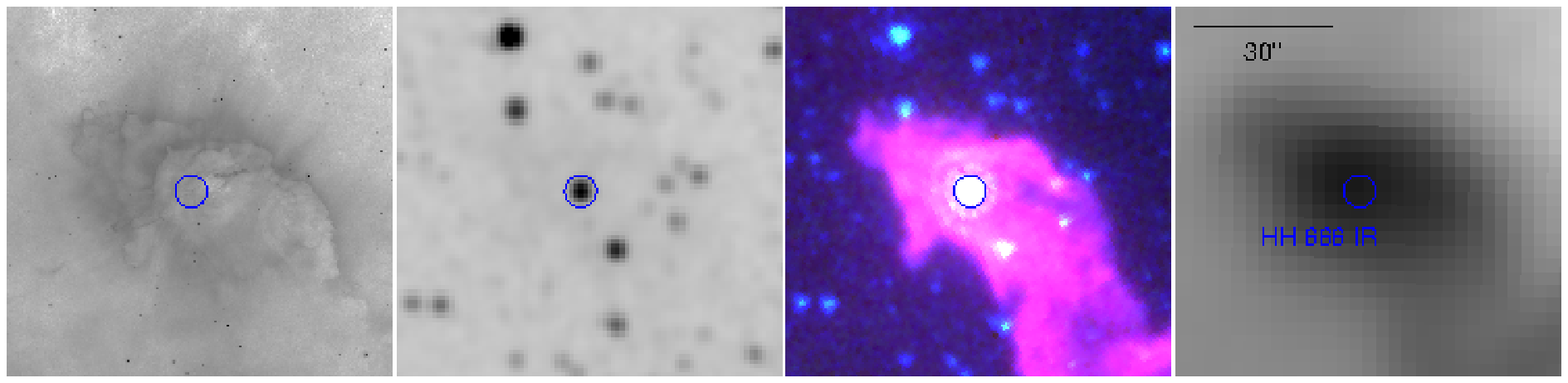}
 \end{minipage}
 \begin{minipage}{\textwidth}
   \includegraphics[width=\textwidth]{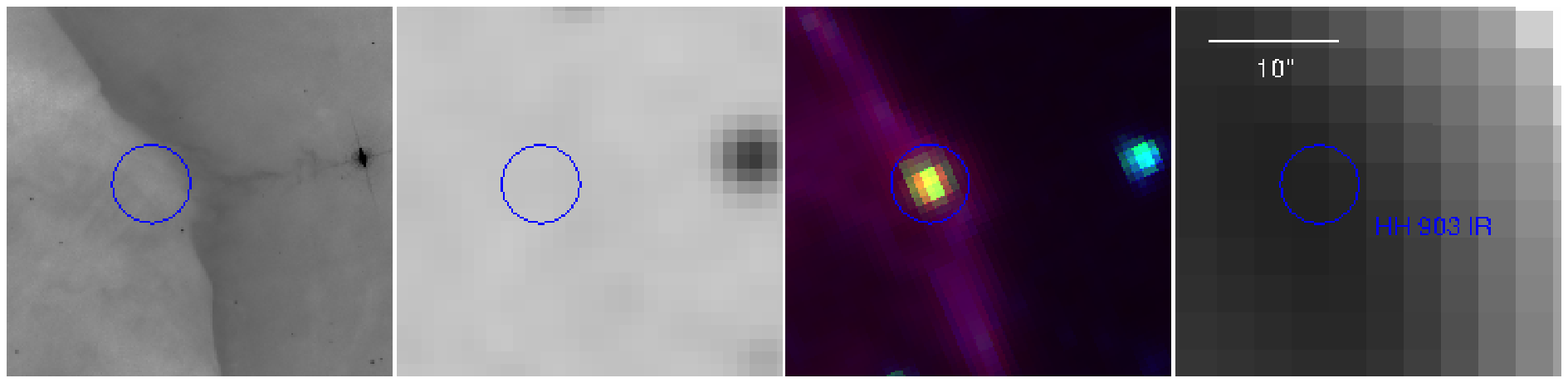}
 \end{minipage}
 \begin{minipage}{\textwidth}
   \includegraphics[width=\textwidth]{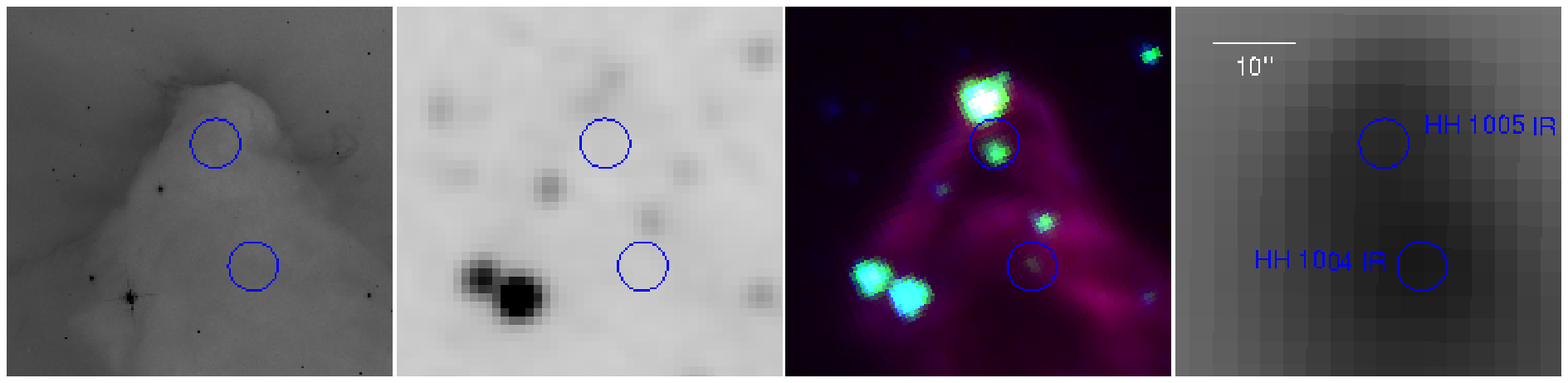}
 \end{minipage}
 \begin{minipage}{\textwidth}
  \includegraphics[width=\textwidth]{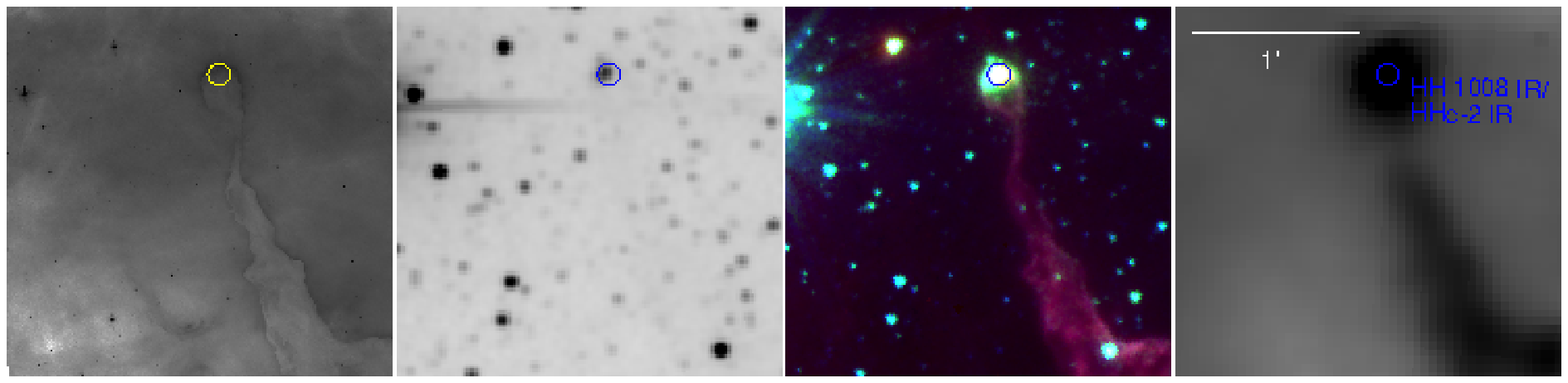}
 \end{minipage} 
 \begin{minipage}{\textwidth}
  \includegraphics[width=\textwidth]{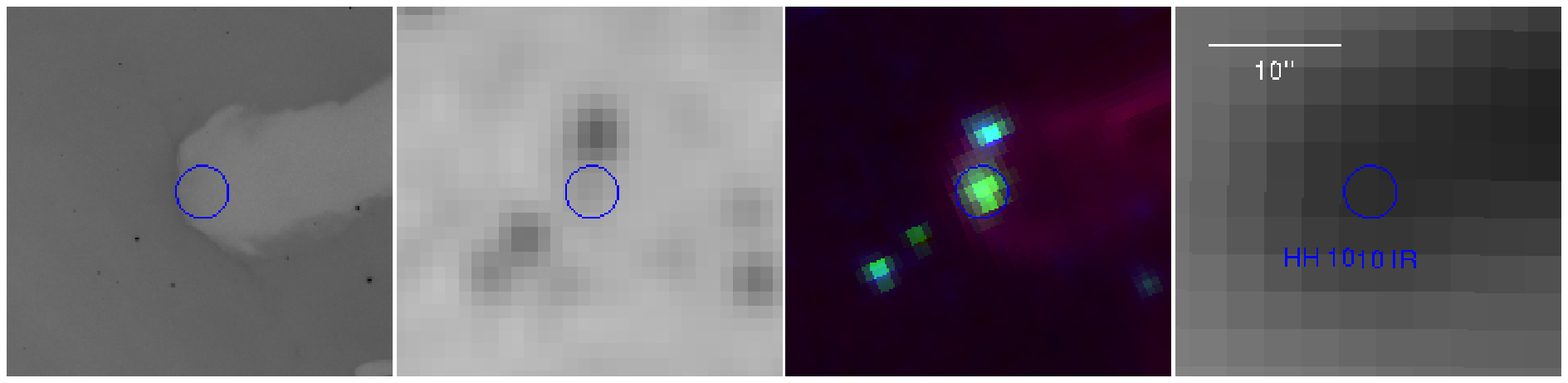}
 \end{minipage} 
 \caption{The Herbig-Haro jets and Herbig-Haro jet candidates in four different wavelengths. Each row represents from left to right the Hubble Space Telescope H$\upalpha$ \citep[][obtained via the Hubble Legacy Archive]{ford2000} image, the 2MASS K-band image \citep[][obtained via the NASA/IPAC Infrared Science Archive]{skrutskie2006}, a Spitzer RGB image with 3.6\mum\ in blue, 4.5\mum\ in green and 8.0\mum\ in red, and Herschel 170\mum\ in the rightmost panel. The sources marked with blue circles (partly yellow for better visibility) and labelled in the Herschel image are those sources of jet objects that are listed in Table~\ref{tab:spitzerherschel_seds}. Spatial scale was chosen individually and is indicated separately for each row.}
 \label{fig:images_hhs}
\end{figure*}

\begin{figure*}
 \centering
  \ContinuedFloat
  \begin{minipage}{\textwidth}
   \includegraphics[width=\textwidth]{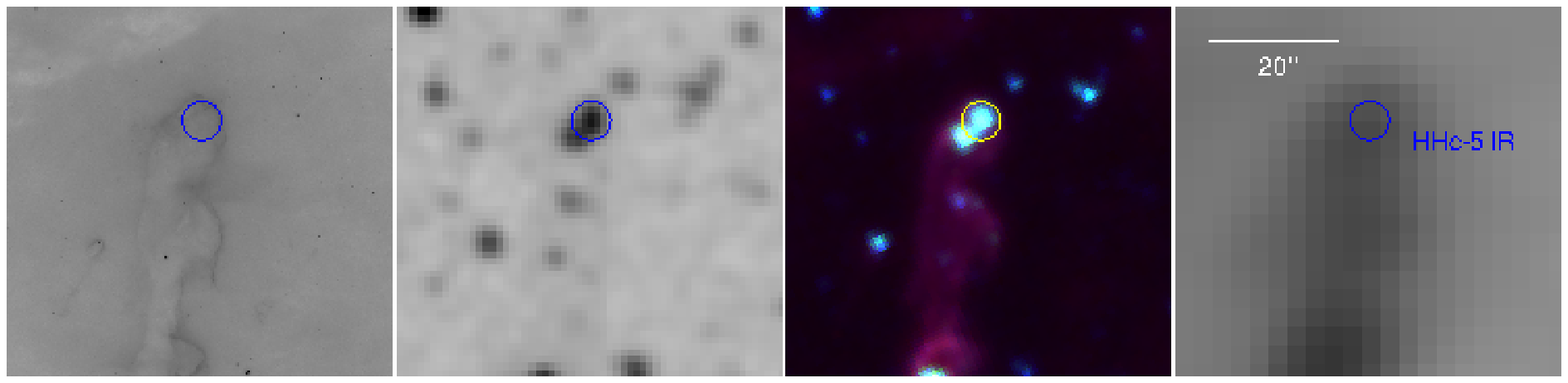}
  \end{minipage}  
 \caption{Continued.}
\end{figure*}

\begin{figure*}
 \centering
 \begin{minipage}{\textwidth}
   \includegraphics[width=\textwidth]{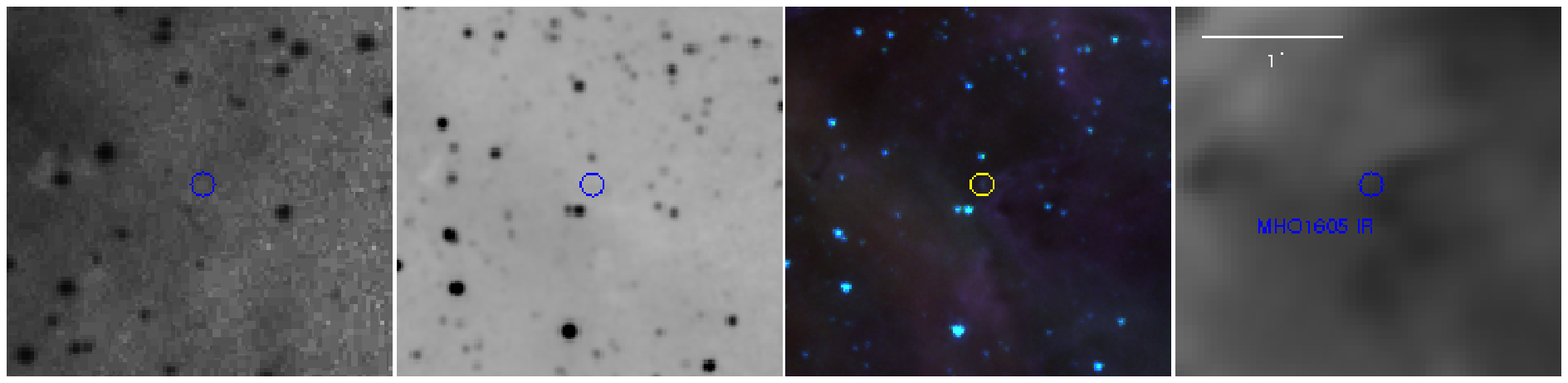}
  \end{minipage}
  \begin{minipage}{\textwidth}
   \includegraphics[width=\textwidth]{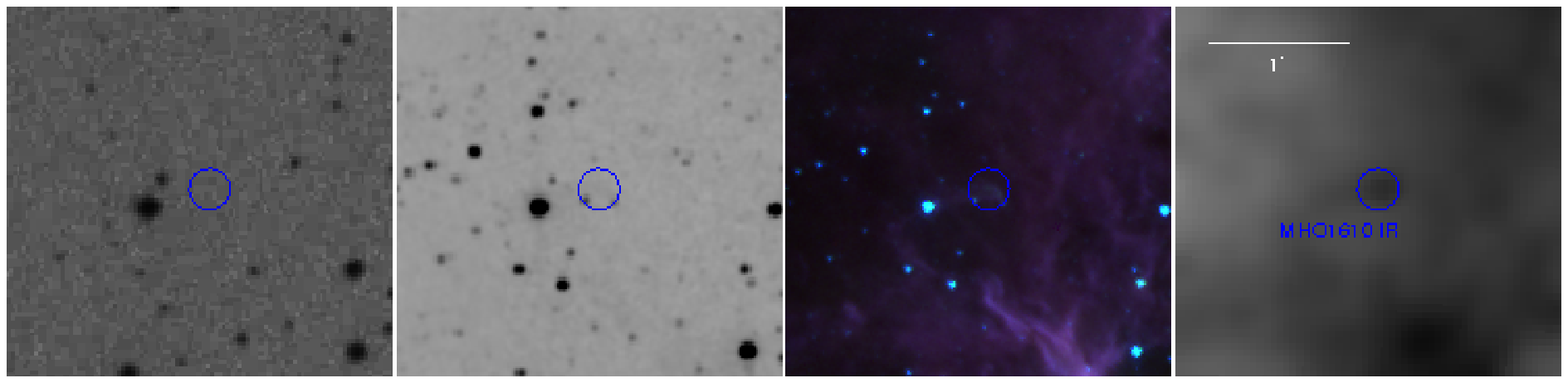}  
  \end{minipage}  
 \caption{As in Fig.~\ref{fig:images_hhs}, but showing the two \mho s identified with Spitzer sources. Instead of the HST image, the leftmost image in each panel shows the STScI Digitized Sky Survey (DSS) optical image.}
 \label{fig:images_mhos}
\end{figure*}
\begin{figure*}
 \centering
 \begin{minipage}{\textwidth}
    \includegraphics[width=\textwidth]{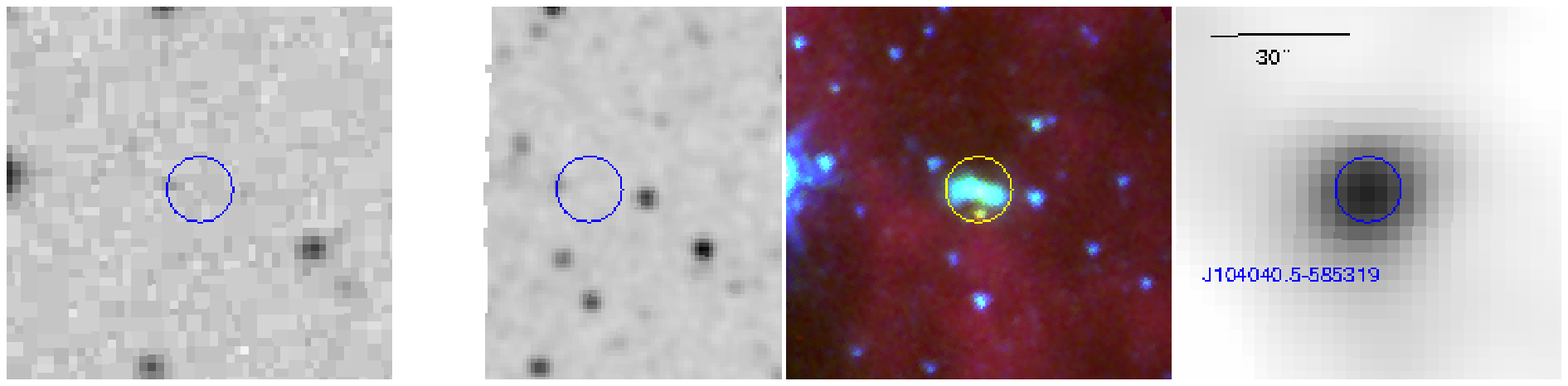}
  \end{minipage}
  \begin{minipage}{\textwidth}
    \includegraphics[width=\textwidth]{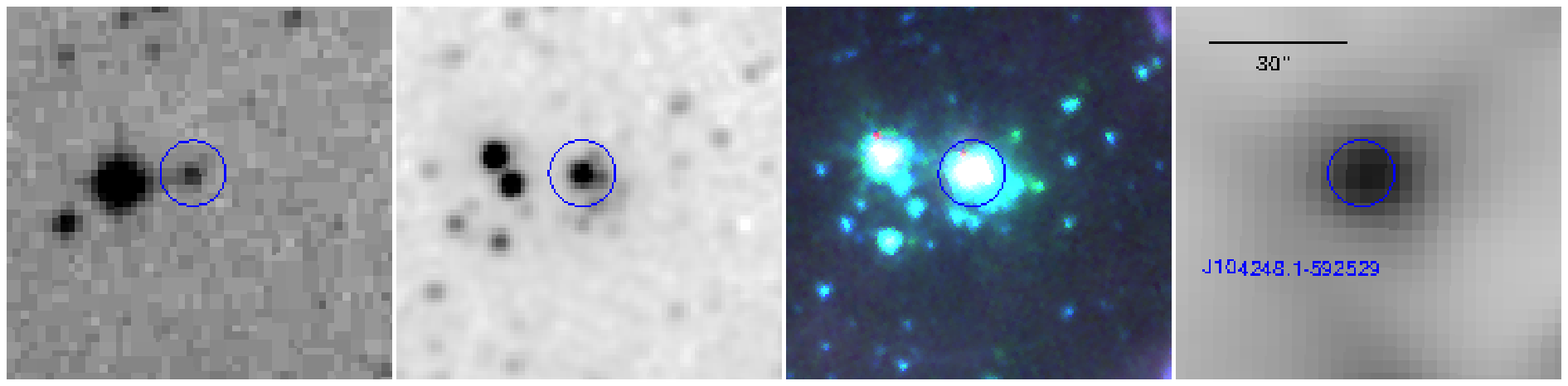}  
  \end{minipage}  
  \begin{minipage}{\textwidth}
    \includegraphics[width=\textwidth]{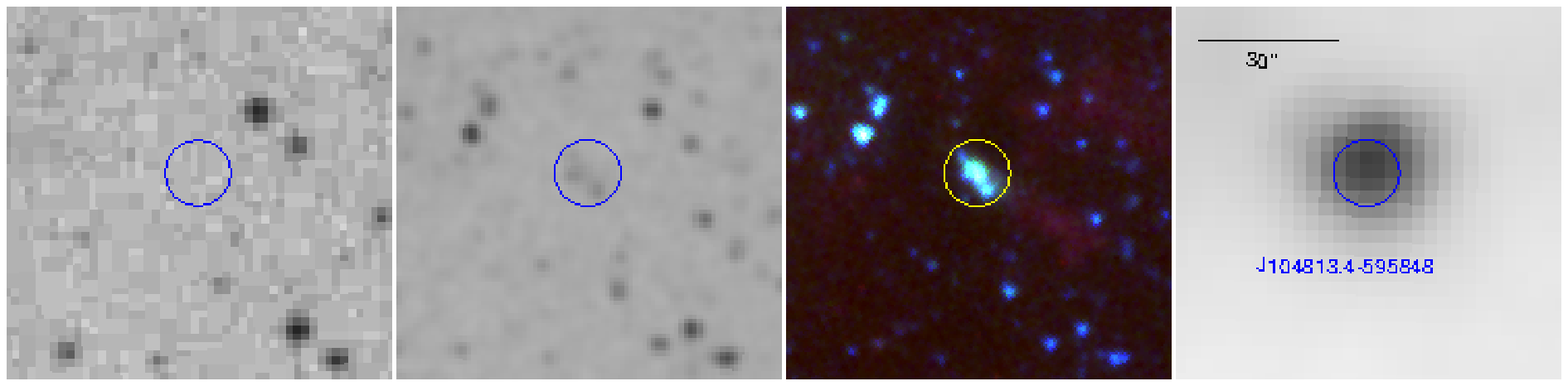}
  \end{minipage}
  \begin{minipage}{\textwidth}
    \includegraphics[width=\textwidth]{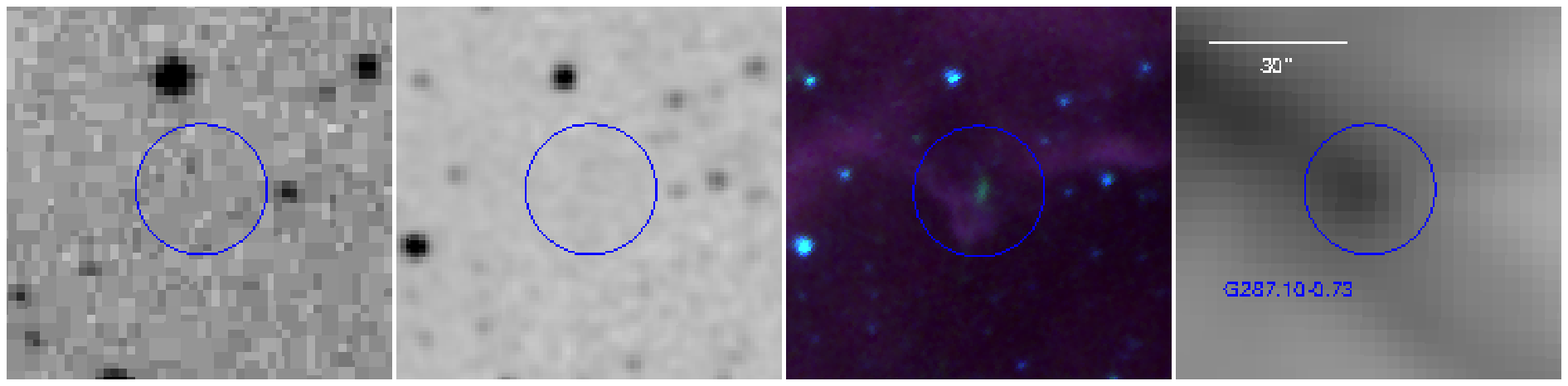}
  \end{minipage}
  \begin{minipage}{\textwidth}
    \includegraphics[width=\textwidth]{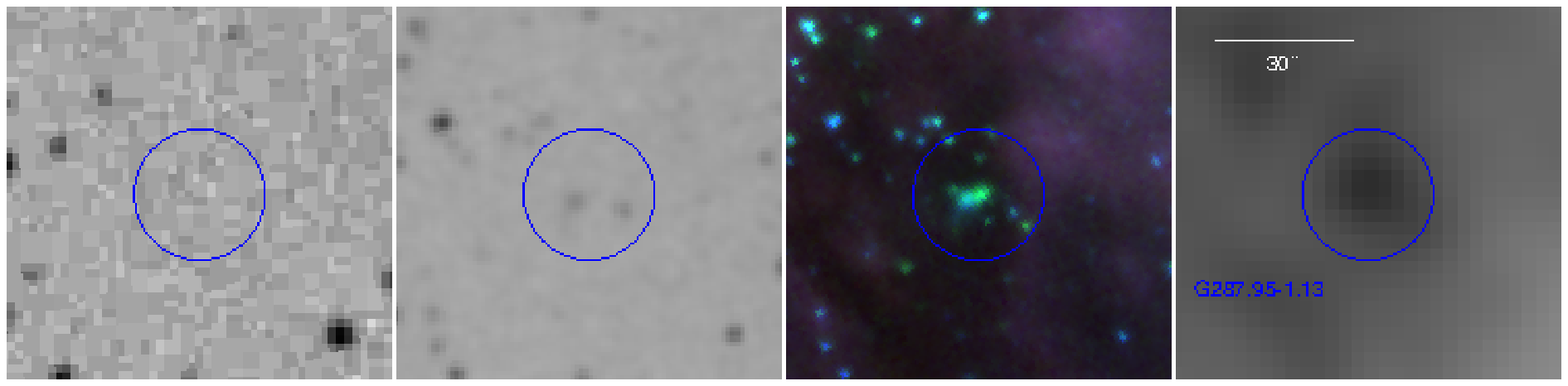}  
  \end{minipage} 
 \caption{As Fig.~\ref{fig:images_mhos}, but showing the seven \ego s in our jet source sample. }
 \label{fig:images_egos}
\end{figure*}

\begin{figure*}
 \centering
  \ContinuedFloat
  \begin{minipage}{\textwidth}
    \includegraphics[width=\textwidth]{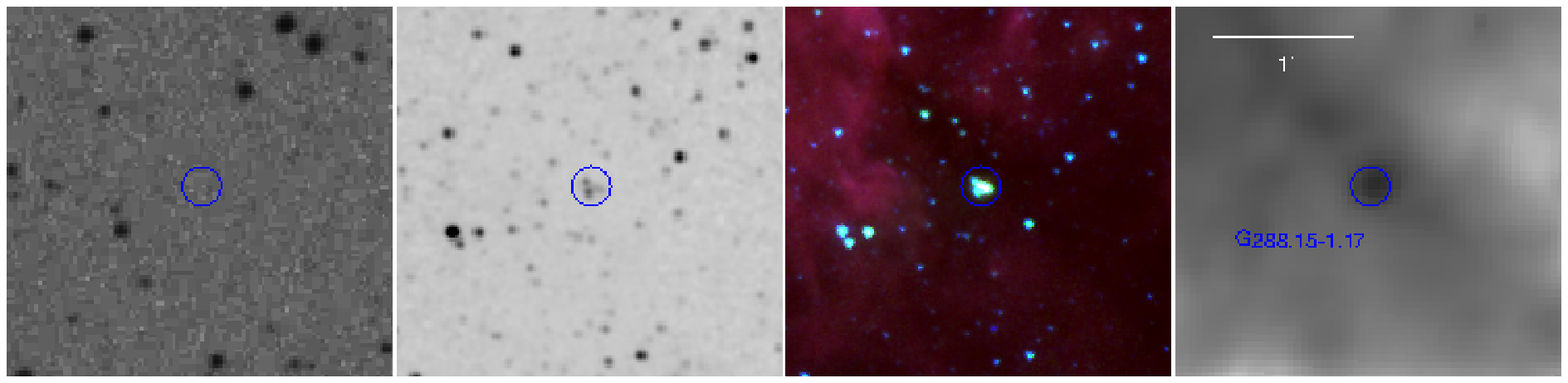}  
  \end{minipage}  
  \begin{minipage}{\textwidth}
    \includegraphics[width=\textwidth]{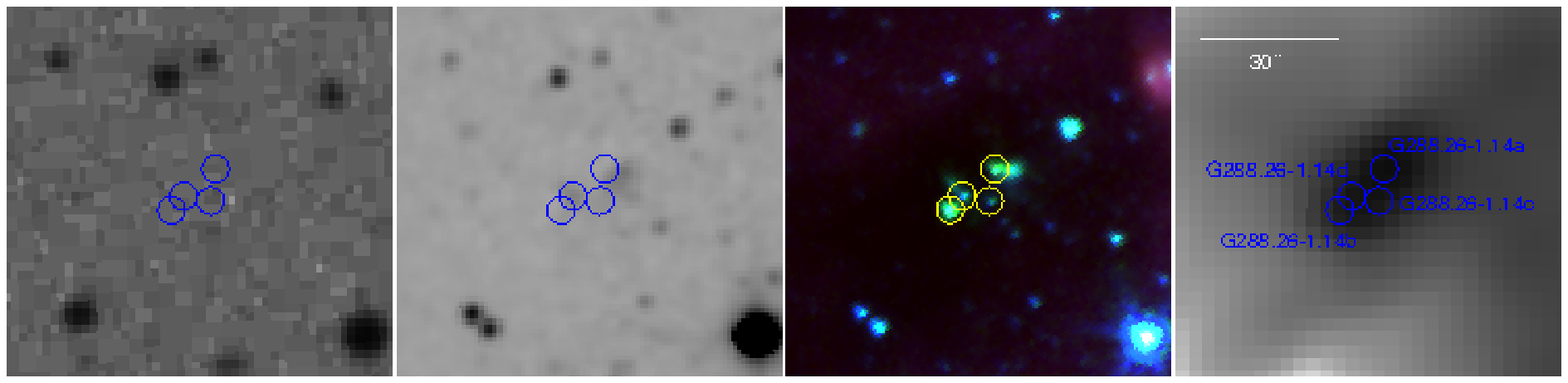}  
  \end{minipage}  
 \caption{Continued.}
\end{figure*}

\begin{figure*}
 \centering
  \begin{minipage}{\textwidth}
    \includegraphics[width=\textwidth]{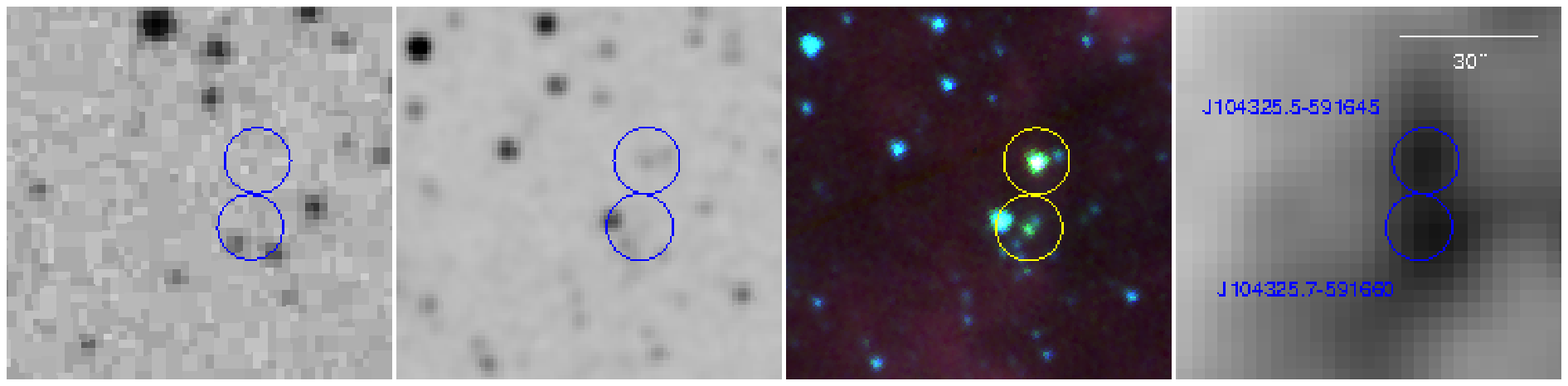}  
  \end{minipage}  
  \begin{minipage}{\textwidth}
    \includegraphics[width=\textwidth]{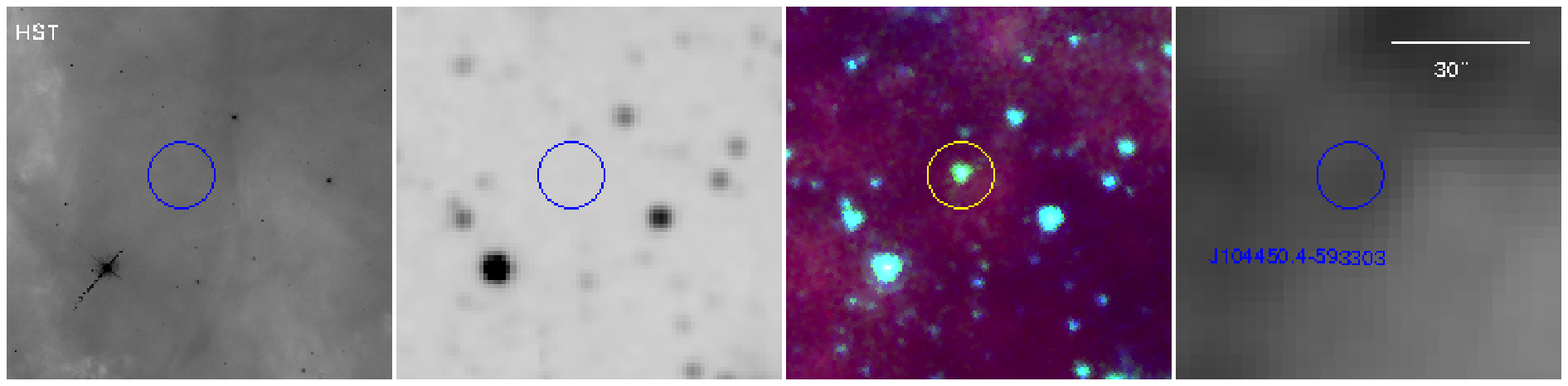}
  \end{minipage}
  \begin{minipage}{\textwidth}
    \includegraphics[width=\textwidth]{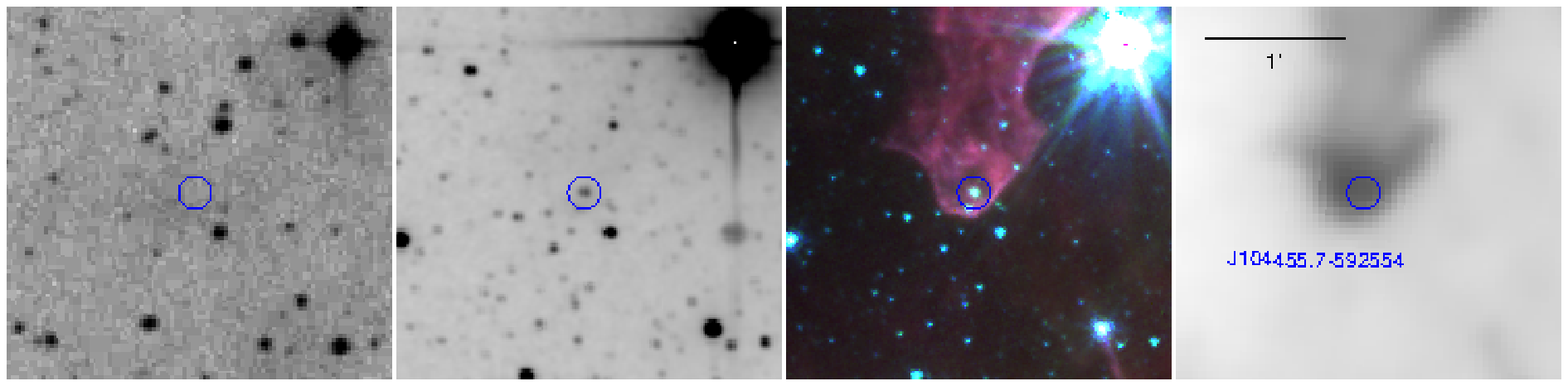}  
  \end{minipage}  
  \begin{minipage}{\textwidth}
    \includegraphics[width=\textwidth]{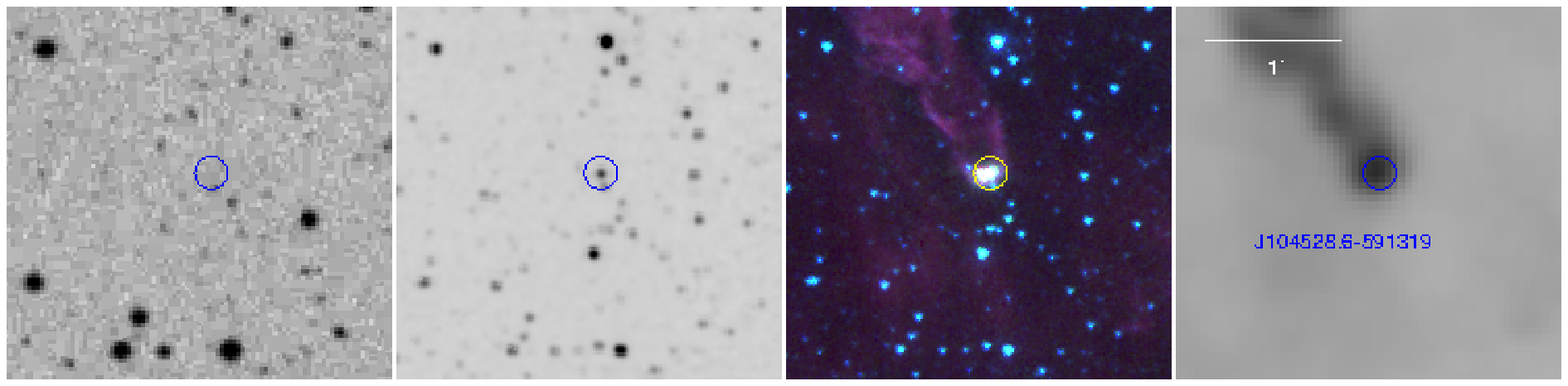}
  \end{minipage}
 \caption{As Fig.~\ref{fig:images_mhos}, showing the five \cgo s in our sample. The leftmost image in each panel is the HST image instead of DSS if indicated by an ``HST'' in the upper left-hand corner.}
 \label{fig:images_gpqs}
\end{figure*}

\subsection{The search for sources of the Herbig-Haro jets}
\label{sec:revealing-hhs}

The mosaic images created from \spitzer\ data as described above were then submitted to visual inspection for probable sources of the 20 Herbig-Haro objects and 15 Herbig-Haro candidates within our field as identified from \hst\ H$\upalpha$ images by \citet[][]{smith2010_jets}.
For 22 \hh\ objects and \hh\ candidates of these 35 objects in our sample, we were able to identify a \spitzer\ point-like source that constitutes a probable jet source, lying on the jet axis or on its projection. Eight of them could also be identified with a point-like \herschel\ source. (Two jets were identified with the same source; see \ref{sec:sources_hh1008hhc2}.)
In one case, that of HH\,900, a \herschel\ source is seen but is probably not associated with a protostar, but rather with the surrounding globule.
The remaining twelve jets have neither \spitzer\ nor \herschel\ sources that correspond with them.
Some outstanding cases are described below.

\subsubsection{HH\,666}

The jet \hh\,666 emanates from out of a pillarhead. This object has already been the subject of a detailed analysis in the infrared by \citet{smith2004_axisofevil}. Prominent in this pillarhead is a very bright \spitzer\ source, which we identified as the source of the jet. It is also associated with a clear point-like \herschel\ source. For the \sed\ fit we added the optical and near-IR fluxes as published by \citet{smith2004_axisofevil} to our \spitzer\ and \herschel\ fluxes.

\subsubsection{HH\,1004 and HH\,1005}

Both the jets assigned as \hh\,1004 and \hh\,1005 by \citet{smith2010_jets} are located towards the tip of the same pillar. Two of the several sources located within this pillar seem to be likely sources for them. The lower part of the clearly two-part structure that is \hh\,1004 is in line with a faint \spitzer\ point-like source that is associated with a distinct compact \herschel\ source and a \hawki\ point-like source. A second \herschel\ source towards the north of it is coincident with a brighter \spitzer\ source and might be a likely source for the jet \hh\,1005.

\subsubsection{HH\,1008 and HH\,c-2}
\label{sec:sources_hh1008hhc2}

\hh\,c-2 is placed right at the top of a thin, elongated pillar, with a very bright \spitzer\ source and a distinct \herschel\ counterpart located inside the pillar head, coincident with a 2MASS point-like source.\footnote{\citet{smith2010_jets} in their Table~3 give coordinates for \hh\,c-2 that are placed within NGC\,3324 and identical with those of the jet \textit{\hh\,c-1 (3324)} and thus are clearly erroneous. However, from their Fig.~10 we were able to identify the jet.} \hh\,1008 is bow-shaped and placed to the south-east of the same pillar head. It is probable that it was caused by the same bright source, since it is well aligned with it, and there is no other candidate source in the vicinity.

\subsubsection{HH\,c-10}

\hh\,c-10 is a bow-like structure and, like \hh\,1008, might have been caused by a source in the nearby head of the pillar close to which it is located. Within this pillarhead, however, there are three distinct \spitzer\ sources while no definite \herschel\ detection can be made. Thus, it remains unclear which, if any, of the \spitzer\ objects might be the source of the Herbig-Haro candidate.

\subsection{The search for sources of the Molecular Hydrogen Emission Objects}
\label{sec:revealing-mhos}

A very similar process was applied to the \mho s detected in \citet{preibisch2011_hawki} which were subjected to visual inspection to search for probable \spitzer\ point-like sources. Four of the six \mho s are located in regions that are comparatively free of \spitzer-detected point-like sources and show no \spitzer\ counterparts themselves. Two of them, \mho s 1605 and 1610, however, show a counterpart within the source catalogue. These are the two objects from the sample of six that are located towards the edge of a pillar or globule, both close to or within known clusters (Tr\,14 and 16).

The \hawki\ data cover only a subfield of the one surveyed in this work, so there might be other H\textsubscript{2} sources that could not be detected. For the region surveyed with \hawki, since all peculiar features in H\textsubscript{2} emission were described in the study in \citet{preibisch2011_hawki}, no further coincidental detections can be reported.

\subsection{The search for extended green objects and compact green objects}
\label{sec:revealing-egosgpqs}

Successively, in the \irac\ images we also visually searched for sources with excessive 4.5\mum-band emission, either extended (\ego s) or point-like (\cgo s). The \rgb\ images (3.6\mum\ in blue, 4.5\mum\ in green and 8.0\mum\ in red) were scaled freely to obtain the best possible visibility of unusually green objects. 
However, \citet{debuizer2005} make the valid point that to be able to identify \ego s, the 4.5\mum\ band usually needs to be scaled up unproportionally high with regard to the red and blue bands. Thus, the weaker emission on the flanks of a relatively extended, bright source might appear to be green only because the scaling was chosen in this way. This is a caveat to be kept in mind when dealing with \ego s surrounding regular point-like sources. Since our \ego s are all well away from bright point-like sources, it is still possible excessively high scaling of the 3.6\mum\ band caused them to appear, but care was taken to make sure they were real phenomena.

Care was also taken to check whether especially \cgo s were not simply imaging faults and green excess was not caused by faulty pixels in the other bands. It can, however, not be excluded that a minority of the peculiarly \cgo s are seen for those reasons and are not real objects. They were selected by visually scanning the processed \spitzer\ images for sources that appeared to the eye to have an excess in the 4.5\mum\ (green) band compared to their immediate surroundings. Overall, we found three \ego s and five \cgo s within the studied area (Fig.~\ref{fig:images_egos}).

\subsubsection{EGOs}

All three new \ego s can clearly be identified with a compact \herschel\ counterpart. Though by nature they are extended, for all three of them we can identify a probable \spitzer\ source for which photometry could be obtained. One of them, J104248.1-592529, is also coincident with a 2MASS source, which allowed for photometry in another three bands.

In addition to the three newly discovered objects we included in our study the four \ego s within the \cnc\ identified by \citet{smith2010_spitzer}. They were subjected to the same analysis as the objects identified through the processes described above. The object described by \citeauthor{smith2010_spitzer} as G288.26-1.14 appears to be a multiple object.
It was therefore split into four objects marked \textit{a} to \textit{d} for the analysis. \cite{smith2010_spitzer} note the multiplicity and point out that some of the green objects in this cluster might not be true \ego s, they should thus be treated with caution. Only for the object designated G288.26-1.14a a compact \herschel\ source is seen, for the other three components no \herschel\ fluxes could be derived.

Likewise, G288.15-1.17 is a composite object. In the \spitzer\ image, it can be recognised as consisting of at least three separate sources (cf.~Fig.~\ref{fig:images_egos}), which is also true for the 2MASS image. Thus, though we could identify \spitzer\ sources and also a corresponding compact \herschel\ object, we did not use the photometry for \sed\ fitting, since it would be highly likely to be comprised of different sources.

For the other two objects we were each able to identify a compact \herschel\ counterpart.

\subsubsection{Compact green objects}

It is the very nature of compact green objects to be identified with a single \spitzer\ source. Four of them also show a compact \herschel\ counterpart. One, J104450.4-593303, is within a region of diffuse \herschel\ emission, but this is not point-like and could thus not be employed for photometry.
For J104528.6-591319, a 2MASS source could be identified and photometry employed for the fitting process.

\section{Characteristics of jet sources}
\label{sec:sources}

\subsection{Radiative transfer modelling of the \sed s}
\label{sec:seds}

From the collected data at \spitzer\ and \herschel\ wavelengths, we consecutively constructed \sed s for the four \spitzer\ \irac\ and five \herschel\ \pacs\ and \spire\ wavelengths, complemented by the \laboca, \hawki, and 2MASS fluxes, as far as photometry was available for the respective sources. For \sed-fitting we used the online tool of \citet{robitaille2007}. This tool compares the input observational data with 200\,000 \sed\ models for \yso s that were precomputed using a 2D radiative transfer code by \citet{whitney2003}. These models have a wide parameter space for the properties of the central object and its environment. The important parameters for our study are stellar mass, circumstellar disk mass, envelope mass, and total luminosity. Because our data sample the peak of the protostar's \sed, they give a good constraint on its total luminosity.

\begin{figure*}
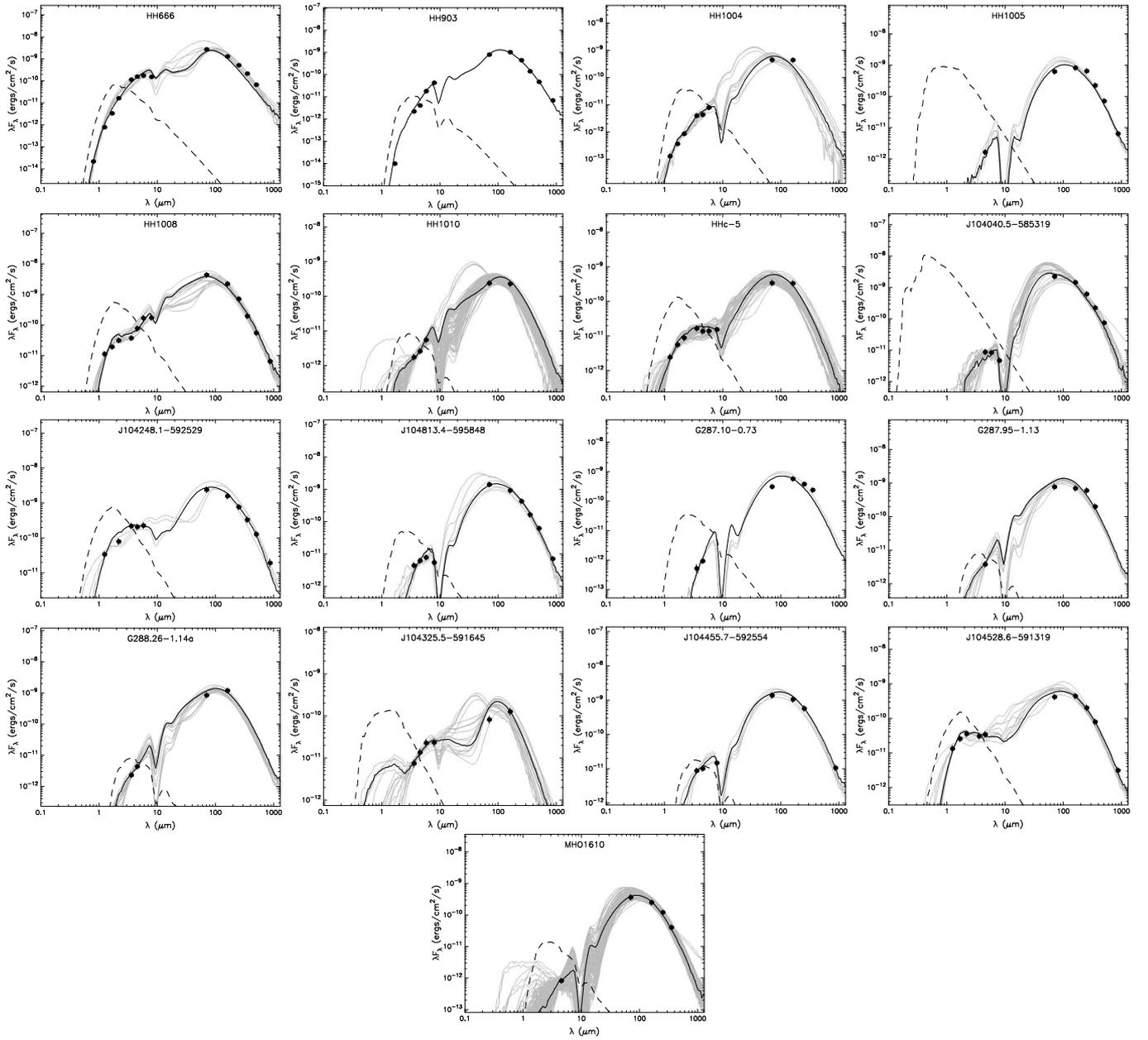

 \centering
 \begin{minipage}{0.24\textwidth}
    \includegraphics[width=\textwidth]{SED_HH666.eps}
  \end{minipage}
  \begin{minipage}{0.24\textwidth}
    \includegraphics[width=\textwidth]{SED_HH903.eps}  
  \end{minipage}  
  \begin{minipage}{0.24\textwidth}
    \includegraphics[width=\textwidth]{SED_HH1004.eps}  
  \end{minipage}  
  \begin{minipage}{0.24\textwidth}
    \includegraphics[width=\textwidth]{SED_HH1005.eps}  
  \end{minipage}  
 \begin{minipage}{0.24\textwidth}
    \includegraphics[width=\textwidth]{SED_HH1008.eps}
  \end{minipage}
  \begin{minipage}{0.24\textwidth}
    \includegraphics[width=\textwidth]{SED_HH1010.eps}  
  \end{minipage}  
 \begin{minipage}{0.24\textwidth}
    \includegraphics[width=\textwidth]{SED_HHc-5.eps}
  \end{minipage} 
  \begin{minipage}{0.24\textwidth}
    \includegraphics[width=\textwidth]{SED_EGO5.eps}  
  \end{minipage}  
  \begin{minipage}{0.24\textwidth}
    \includegraphics[width=\textwidth]{SED_EGO6.eps}
  \end{minipage}
  \begin{minipage}{0.24\textwidth}
    \includegraphics[width=\textwidth]{SED_EGO7.eps}  
  \end{minipage} 
  \begin{minipage}{0.24\textwidth}
    \includegraphics[width=\textwidth]{SED_073.eps}  
  \end{minipage} 
  \begin{minipage}{0.24\textwidth}
    \includegraphics[width=\textwidth]{SED_113.eps}  
  \end{minipage} 
  \begin{minipage}{0.24\textwidth}
    \includegraphics[width=\textwidth]{SED_114a.eps}  
  \end{minipage} 
  \begin{minipage}{0.24\textwidth}
    \includegraphics[width=\textwidth]{SED_GPQ11.eps}  
  \end{minipage} 
  \begin{minipage}{0.24\textwidth}
    \includegraphics[width=\textwidth]{SED_GPQ14.eps}
  \end{minipage}
  \begin{minipage}{0.24\textwidth}
    \includegraphics[width=\textwidth]{SED_GPQ15.eps}
  \end{minipage}
  \begin{minipage}{0.24\textwidth}
    \includegraphics[width=\textwidth]{SED_MHO1610.eps}  
  \end{minipage}  
 \caption{Spectral energy distributions of those 17 objects for which we could determine fluxes in at least four bands overall.  Filled circles mark the input fluxes. The black line shows the best fit, and the grey lines show subsequent good fits. The dashed line represents the stellar photosphere corresponding to the central source of the best-fitting model, as it would appear in the absence of circumstellar dust (but including interstellar extinction).}
 \label{fig:seds}
\end{figure*}

\herschel, \spitzer, and \laboca\ fluxes as listed in Table\,\ref{tab:spitzerherschel_seds}, complemented by \hawki\ and 2MASS fluxes where available, were used for the compilation of the \sed s. No fits were performed for sources with only \spitzer\ fluxes or fewer than four data points overall because no reasonable constraints to the model could be obtained in these cases. 
For the fits, the distance to all objects was assumed to be 2.3\,kpc, and the interstellar extinction range was restricted to to $A_V = 0 \dots 40$\,mag. We assumed an uncertainty of 20\% for all fluxes.  In addition to the best-fit model, we show the range of possible parameters that can be derived from models within the range of $\chi^2\!/\nu - \chi_\mathrm{best}^2/\nu < 2$ (with $\nu$ representing the number of data points). These models are shown as grey lines in the plots in Fig.~\ref{fig:seds}. The resulting model parameters are listed in Table\,\ref{tab:modelpars}. It gives the best-fit value together with the range constrained by the above $\chi^2$ criterion. The resulting \sed s are shown in Fig.\,\ref{fig:seds}.

Notably, all the sources in our sample are well modelled by models possessing a circumstellar disk. 
The masses of the jet-emitting protostars are within a range of $~ 1 - 10\,M_\odot$.
Thus, they all are of low masses, with luminosities of the best-fit models varying between 40 and $1600\,L_\odot$. Most of the luminosities are in the lower range below a few hundred $L_\odot$, with three notable exceptions that exhibit luminosities of more than $700\,L_\odot$.
Sources of \cgo s seem to be lower-mass (best-fit values $\sim 1-5\,M_\odot$) than those of \ego s (best-fit values $\sim 5-8\,M_\odot$), while those objects emitting \hh\ jets are of varying masses (best-fit values $\sim 2-7\,M_\odot$). Envelope masses of \ego\ sources are noticeably higher than most of those for other jet objects. All best-fit envelope masses for \ego\ sources are at least $350\,M_\odot$, while many of those for the other objects (6 out of 11, 55\%) are far below $100\,M_\odot$. 
However, five of those also range far above that value. There is no distinct pattern recognisable for the disk masses, if any, that \hh\ jet sources have slightly higher disk masses than those of the other jet objects. 
Still, the smallness of the sample means that these observations of course only have limited statistical value. For \mho s especially, we only have a single specimen so that no qualitative statements are possible.

\begin{landscape}
\begin{table}
\caption[]{Source fluxes as obtained in the Spitzer (3.6\mum, 4.5\mum, 5.8\mum, and 8.0\mum) and Herschel (70\mum, 160\mum, 250\mum, 350\mum, and 500\mum) bands, complemented with \laboca\ 870\mum\ photometry and JHK\textsubscript{s} photometry as obtained with HAWK-I or from the 2MASS point-source catalogue.}
\label{tab:spitzerherschel_seds}
\centering
\begin{tabular}{llrrrrrrrrrrrrr}
\hline\hline
\noalign{\smallskip}
\multicolumn{2}{c}{\multirow{2}{*}{Jet Source}}  & 3.6\mum  &   4.5\mum  &    5.8\mum  &  8.0\mum & 70\mum & 160\mum & 250\mum & 350\mum & 500\mum & 870\mum  & J & H & K\textsubscript{s} \\
\multicolumn{2}{c}{}  & [mJy]  & [mJy] & [mJy] & [mJy] & [Jy]  & [Jy] & [Jy] & [Jy] & [Jy] & [Jy] & [mJy] & [mJy] & [mJy] \\
\noalign{\smallskip}
\hline
\noalign{\smallskip}
\hh666 IR     & J104351.5--595521 & 137   & 238   & 338  & 405  & 64.6 & 70.5 & 43.9 & 25.6  & 11.4 & --    & 0.33\tablefootmark{a}  & 1.9\tablefootmark{a}   & 12\tablefootmark{a}  \\
\hh902 IR     & J104401.8--593030 & --    & 4.17  & 29.6 & 70.0 & --   & --   & --   & --    & --   & -- & -- & -- & --  \\ 
\hh903 IR     & J104556.4--600608 & 2.68  & 6.24  & 35.4 & 115  & 19.1 & 55.7 & 37.9 & 16.7  & 8.07 & 2.03  & --    & --    & --    \\ 
\hh1004 IR    & J104644.8--601021 & 4.75  & 6.58  & 15.2 & --   & 10.6 & 23.9 & --   & --    & --   & --    & 0.0540 & 0.206 & 0.624 \\ 
\hh1005 IR    & J104644.2--601035 & --    & 2.50  & --   & --   & 14.8 & 44.4 & 55.0 & 26.6  & 12.0 & 1.89  & --    & --    & --    \\ 
\hh1006 IR    & J104632.9--600354 & 2.57  & 3.06  & 10.6 & 23.1 & --   & --   & --   & --    & --   & --    & --    & --    & --    \\ 
\hh1008/c-2 IR& J104445.4--595555 & 44.8  & 117   & 326  & 461  & 103  & 120  & 59.9 & 23.2  & 9.32 & 1.90  & 4.71  & 10.8  & 22.8  \\ 
\hh1010 IR    & J104148.7--594338 & 2.11  & 3.96  & 10.8 & --   & 5.67 & 12.5 & --   & --    & --   & --    & --    & --    & --    \\ 
\hh1012 IR    & J104438.7--593008 & 5.29  & 5.27  & 5.92 & --   & --   &  --  &  --  & --    &  --  & --    & --    & --    & --    \\
\hh1013 IR    & J104419.2--592612 & 8.46  & 11.7  & 16.7 & 25.1 & --   & --   & --   & --    & --   & --    & --    & --    & --    \\ 
\hh1017 IR    & J104441.5--593357 & 4.98  & 5.12  & --   & --   & --   & --   & --   & --    & --   & --    & --    & --    & --    \\
\hh1018 IR    & J104452.9--594526 & 4.54  & 5.39  & --   & 10.5 & --   & --   & --   & --    & --   & --    & --    & --    & --    \\
\hhc1 IR      & J104405.4--592941 & 6.00  & --    & --   & --   & --   & --   & --   & --    & --   & --    & --    & --    & --    \\ 
\hhc3 IR      & J104504.6--600303 & 0.611 & 0.712 & 2.34 & 5.02 & --   & --   & --   & --    & --   & --    & --    & --    & --    \\ 
\hhc5 IR      & J104509.4--600203 & 20.0  & 21.1  & 27.9 & 41.2 & 8.12 & 18.1 & --   & --    & --   & --    & 1.02  & 3.17  & 6.52  \\ 
\hhc6 IR      & J104509.2--600220 & --    & 1.29  & --   & --   & --   & --   & --   & --    & --   & --    & --    & --    & --    \\ 
\hhc7 IR      & J104513.0--600259 & 2.25  & 2.00  & --   & --   & --   & --   & --   & --    & --   & --    & --    & --    & --    \\ 
\hhc8 IR      & J104512.0--600310 & 5.48  & 8.33  & 13.2 & 16.0 & --   & --   & --   & --    & --   & --    & --    & --    & --    \\ 
\hhc13 IR     & J104400.6--595427 & 13.2  & 12.5  & 13.8 & 17.1 & --   & --   & --   & --    & --   & --    & --    & --    & --    \\ 
\hhc15 IR     & J104353.9--593245 & 274   & 285   & 283  & 239  & --   & --   & --   & --    & --   & --    & --    & --    & --    \\ 
\ego          & J104040.5--585319 & --    & 13.4  & 16.3 & 12.6 & 53.2 & 79.4 & 52.7 & 26.8  & 12.7 & --    & --    & --    & --    \\ 
\ego          & J104248.1--592529 & 265   & 318   & 440  & --   & 57.3 & 85.9 & 65.4 & 39.4  & 21.8 & 5.73  & 14.0  & --    & 58.1  \\ 
\ego          & J104813.4--595848 & 5.34  & 9.52  & 15.2 & 14.6 & 33.4 & 49.8 & 36.2 & 19.6  & 10.5 & 2.09  & --    & --    & --    \\
G287.10-0.73  & J104114.3--593239 & 0.645 & 1.42  & --   & --   & 7.36 & 31.0 & 32.2 & 28.3  & --   & --    & --    & --    & --    \\
G287.95-1.13  & J104542.0--601732 & --    & 5.80  & --   & --   & 18.4 & 38.2 & 51.0 & 23.8  & --   & --    & --    & --    & --    \\
G288.26-1.14a & J104750.3--602618 & 2.83  & 6.70  & --   & --   & 20.0 & 65.0 & --   & --    & --   & --    & --    & --    & --    \\
G288.26-1.14b & J104752.1--602627 & 4.02  & 14.5  & 28.8 & 31.3 & --   & --   & --   & --    & --   & --    & --    & --    & --    \\
G288.26-1.14c & J104750.9--602625 & 1.13  & 1.93  & --   & --   & --   & --   & --   & --    & -    & --    & --    & --    & --    \\
G288.26-1.14d & J104751.7--602624 & 2.25  & --    & --   & --   & --   & --   & --   & --    & --   & --    & --    & --    & --    \\
\cgo          & J104325.5--591645 & 8.99  & 20.8  & 44.5 & 63.0 & 1.95 & 6.93 & --   & --    & --   & --    & --    & --    & --    \\ 
\cgo          & J104325.7--591660 & 1.60  & 3.26  & 6.69 & 8.22 & --   & --   & 10.5 & 7.10  & 4.78 & --    & --    & --    & --    \\ 
\cgo          & J104450.4--593303 & 3.18  & 21.3  & 9.17 & 12.0 & --   & --   & --   & --    & --   & --    & --    & --    & --    \\ 
\cgo          & J104455.7--592554 & 10.6  & 15.4  & --   & 39.6 & 32.7 & 57.4 & 48.8 & --    & --   & 3.11  & 0.821 & 1.64  & 2.75  \\ 
\cgo          & J104528.6--591319 & 37.1  & 52.7  & --   & --   & 9.96 & 24.1 & 17.2 & 9.46  & --   & 0.923 & 5.63  & 14.6  & 26.1  \\ 
MHO1605 IR    & J104351.5--593911 & 1.27  & 1.79  & 8.19 & 17.9  & --  & --   & --   & --    & --   & --    & --    & --    & --    \\
MHO1610 IR    & J104608.3--594525 & --    & 1.26  &   -- &   -- & 8.68 & 13.6 & 10.3 & 4.80  & --    & --   & --    & --    & --    \\
\hline
\end{tabular}
\tablefoot{The first column indicates the type of the associated jet object: \ego\ for extended green objects, \cgo\ for green point sources. The \ego s from \citet{smith2010_spitzer} bear the identifiers given there. For sources of Herbig-Haro objects and Herbig-Haro candidates the identifier of the corresponding jet object following \citet{smith2010_jets} is given, for the \mho s their catalogue number as assigned by \citet{preibisch2011_hawki}. For all sources their coordinates are given in the second column. \tablefoottext{a}{For HH 666 IR we adopted the JHK\textsubscript{s} fluxes measured by \citet{smith2004_axisofevil} with SOFI.}}
\end{table}
\end{landscape}

\begin{table*}
\caption[]{Model parameters for the jet sources as obtained from the \citet{robitaille2007} models.}
\label{tab:modelpars}
\centering
\begin{tabular}{c r r r r r r r r c}
\hline\hline
\noalign{\smallskip}
\multirow{2}{*}{Jet Source} & \multicolumn{2}{c}{Stellar Mass} & \multicolumn{2}{c}{Disk Mass}   & \multicolumn{2}{c}{Envelope Mass} & \multicolumn{2}{c}{Total Luminosity} & Best-Fit \\
                            & \multicolumn{2}{c}{$[M_\odot]$}  & \multicolumn{2}{c}{$[M_\odot]$} & \multicolumn{2}{c}{$[M_\odot]$}   & \multicolumn{2}{c}{$[L_\odot]$} & Model \\
\noalign{\smallskip}
\hline
\noalign{\smallskip}
J104351.5--595521 & 3.2 & [1.7 -- 9.2] &  1.26\timestento{-1}    &   [1.93\timestento{-2} -- 5.68\timestento{-1}]    &   20	 &    [5.3 -- 410]  &  420   &   [390 -- 1400] & 3014036 \\
J104556.4--600608 & 4.3 & [4.3 -- 4.3] &  1.74\timestento{-2}    &   [1.74\timestento{-2} -- 1.74\timestento{-2}]    &   490     &    [490 -- 490]  &  250   &   [250 --  250] & 3018560 \\
J104644.8--601021 & 4.6 & [2.6 -- 7.0] &  1.40\timestento{-2}    &   [7.32\timestento{-3} -- 2.26\timestento{-1}]    &   20	 &    [4.9 -- 250]  &  150   &   [81 -- 1600]  & 3016324 \\
J104644.2--601035 & 5.5 & [5.5 -- 7.3] &  2.07\timestento{-1}    &   [4.25\timestento{-3} -- 5.87\timestento{-1}]    &   760     &    [480 -- 760]  &  230   &   [230 --  440] & 3013239 \\
J104445.4--595555 & 7.0 & [2.1 -- 8.1] &  1.30\timestento{-1}    &   [1.60\timestento{-3} -- 3.42\timestento{-1}]    &   560     &    [ 94 -- 640]  &  710   &   [370 -- 1000] & 3006553 \\
J104148.7--594338 & 1.9 & [1.3 -- 6.7] &  2.08\timestento{-2}    &   [7.76\timestento{-4} -- 2.80\timestento{-1}]    &   77	 &    [6.1 -- 250]  &	64   &   [40 -- 1400]  & 3008354 \\
J104509.4--600203 & 4.3 & [1.4 -- 6.4] &  4.79\timestento{-3}    &   [2.10\timestento{-4} -- 2.73\timestento{-1}]    &   43	 &    [3.9 -- 350]  &  130   &   [56 --  310]  & 3019527 \\
J104040.5--585319 & 8.3 & [1.9 -- 10 ] &  1.81\timestento{-2}    &   [2.73\timestento{-3} -- 6.71\timestento{-1}]    &   350     &    [ 39 -- 1300] & 1700   &   [300 -- 3400] & 3002514 \\
J104248.1--592529 & 8.1 & [5.0 -- 8.3] &  7.32\timestento{-3}    &   [7.32\timestento{-3} -- 5.06\timestento{-1}]    &   1500    &    [200 -- 1700] &  820   &   [460 -- 1000] & 3006025 \\
J104813.4--595848 & 6.6 & [3.6 -- 8.3] &  5.87\timestento{-1}    &   [2.77\timestento{-3} -- 5.87\timestento{-1}]    &   550     &    [250 -- 750]  &  330   &   [220 -- 2500] & 3013941 \\
J104114.3--593239 & 6.5 & [5.5 -- 6.5] &  1.91\timestento{-3}    &   [1.91\timestento{-3} -- 2.07\timestento{-1}]    &   430     &    [430 -- 760]  &  250   &   [230 --  250] & 3020073 \\
J104542.0--601732 & 4.9 & [3.0 -- 7.3] &  7.72\timestento{-2}    &   [2.77\timestento{-3} -- 5.87\timestento{-1}]    &   540     &    [300 -- 550]  &  250   &   [190 --  420] & 3014735 \\
J104750.3--602618 & 4.9 & [3.0 -- 6.6] &  7.72\timestento{-2}    &   [2.77\timestento{-3} -- 5.87\timestento{-1}]    &   540     &    [ 66 -- 550]  &  250   &   [190 --  380] & 3014735 \\
J104325.5--591645 & 1.3 & [0.8 -- 5.7] &  9.63\timestento{-3}    &   [6.35\timestento{-5} -- 1.86\timestento{-1}]    &   10	 &    [1.9 --  53]  &	38   &   [23 --  760]  & 3002569 \\
J104455.7--592554 & 4.6 & [4.6 -- 7.7] &  8.23\timestento{-2}    &   [4.79\timestento{-2} -- 5.87\timestento{-1}]    &   1200    &    [550 -- 1200] &  310   &   [230 --  570] & 3000552 \\
J104528.6--591319 & 2.6 & [1.5 -- 4.8] &  4.25\timestento{-2}    &   [1.05\timestento{-3} -- 2.16\timestento{-1}]    &   270     &    [ 65 -- 300]  &  100   &   [94 --  200]  & 3003259 \\
J104608.3--594525 & 3.6 & [1.4 -- 6.4] &  1.34\timestento{-3}    &   [2.52\timestento{-4} -- 2.71\timestento{-1}]    &   80	 &    [9.3 -- 250]  &  110   &   [52 --  260]  & 3004022 \\
\hline		  
\end{tabular}	  
\tablefoot{For each model parameter, the best-fit value is given in the respective first column, followed by a range defined by the minimum and maximum values obtained from models constrained by a $\chi^2$ criterion (see Section \ref{sec:seds}). The last column gives the identifier of the best-fit model.}
\end{table*}

\section{Discussion}
\label{sec:discussion}

We surveyed 55 jet objects and found corresponding point-like sources for 36 of them.
For three of the \cgo s and seven of the \ego s we were able to derive the \sed s of newly identified IR sources. 
Out of 35 \hh\ jets and \hh\ jet candidates, we could identify sources for 22 of them and model \sed s for seven. Out of the six \mho s within the \hawki\ field -- which is smaller than the entire field surveyed here -- we could only identify sources for two and find an \sed\ model for one.

It is important to note that this study is not able to identify all the jet objects in the \cnc. The majority of jets are just too faint and fall below the sensitivity limits of our data. Consequently, out of the members of the \cnc\ that are currently undergoing jet-emitting phases, we can only observe a low percentage of the most luminous ones. This explains why out of the very large number of young stars within the surveyed region only so few jets are detected.

The stellar-mass estimates derived from \sed\ fitting are in the range $\sim 1\,M_\odot$ to $\sim 10\,M_\odot$. This range agrees very well with the expectations in the following sense: The lower mass limit of $\sim 1\,M_\odot$ is probably a consequence of the detection limit. Although the \spitzer\ images are deep enough to detect stars below this mass limit, one has to keep in mind that it is necessary to detect emission from the jet, too. That the mass flow rates of jets and outflows from \yso s are roughly correlated to the stellar luminosity \citep[and thus also to the stellar mass; \eg][]{reipurth2001} explains why we do not detect jets from YSOs of lower masses.
The upper mass limit is more interesting, since there is no reason why the jets from high-mass protostars should not be observable, if such jets were present. The absence of jet-driving high-mass ($M \ga 20\,M_\odot$) protostars therefore suggests that no such objects are present. This agrees with the conclusion derived from the analysis of the sub-mm data of the cloud structure by \citet{preibisch2011_laboca} that the clouds present today in the Carina Nebula are not dense and massive enough for high-mass star formation. This suggests that the currently ongoing star formation process in the \cnc\ is qualitatively different from the processes that led to the formation of dozens of very high-mass stars \citep[$M \ga 50\,M_\odot$, \eg][]{smith2006} in the \cnc\ a few million years ago.

Further conclusions about the star-formation process can be drawn from the properties of the jets and the jet-driving protostars. The first interesting result is found by comparing the number of jets detected in the infrared images to the number of HH jets detected in the optical HST images. For a quantitative analysis, we consider the $\approx 1.0\degr \times 1.3\degr$ region investigated by \citet[][their Fig.~1]{smith2010_jets}. Their (mostly non-contiguous) HST pointings cover $\approx 15\%$ of this area and have led to the detection of 34 HH jets and jet-candidates. If the full area had been observed by HST, the number of detected jets in this area would very likely be larger, perhaps\footnote{We have to consider that the positions of the individual \hst\ pointings were targeted on particularly promising locations and the true spatial distribution of HH jets is presumably not perfectly homogeneous; thus, the number of jets is presumably smaller than an estimate based simply on the area-filling factor of $\approx 15\%$.} $\sim 70$.

The number of jets detected in the infrared images of the same area is 20 (9 EGOs, 5 CGO, and 6 MHOs), \ie\ substantially smaller. We are confident that this is not an effect of the cloud extinction, because the sub-mm observations \citep{preibisch2011_laboca} have shown that the column densities and the extinction of almost all clouds in the region are rather low, generally $A_V \la 15$~mags. The clouds are therefore largely transparent for light in the \spitzer\ bands, so that we do not expect to miss a significant number of jets due to cloud extinction.

For comparing the number of optical \hh\ jets to the infrared jets one has to keep in mind that these two groups are intrinsically different: The optical \hh\ objects trace jets moving in the atomic gas outside the molecular clouds
and originate in protostars located close to the surfaces of clouds. The infrared jets, on the other hand, trace molecular material inside the clouds \citep[\eg][]{elias1980}. Since almost all clouds should be essentially transparent at \spitzer\ wavelengths, one expects to see considerably more infrared jets than optical jets, assuming that the jet-driving sources are distributed in a spatially homogeneous way throughout the entire volume of the clouds. In many observations of nearby star-forming clouds, this expectation is confirmed \citep[see \eg][for the case of the NGC~1333 cloud]{gutermuth2008}.

In the case of the \cnc, however, we see the opposite trend, \ie\ the number of infrared jets 
is smaller than the number of optical jets. This suggests that most jet-driving protostars are located at the surfaces
of the clouds, rather than in the inner cloud regions.

\section{Conclusions}
\label{sec:conclusions}

Our search for jets in wide-field \spitzer\ \irac\ mosaic images of the \cnc\ led to the detection of three new \ego s and five \cgo s in and immediately around the area studied previously with the \hst\ by \citet{smith2010_jets}. We also identified sources for 23 of the 34 HH jets in this area.
Combining \spitzer\ and \herschel\ photometry with complementary \laboca\ and \hawki/2MASS data, we obtained the spectral energy distributions of the jet sources through radiative transfer modelling \citep{robitaille2006, robitaille2007} to estimate basic stellar and circumstellar parameters for 17 sources overall, where photometry in at least four bands could be obtained.

We find that the jet-driving protostars generally have low to intermediate masses ($\sim1-10\,M_\odot$), which corroborates the notion that there is no current formation of high-mass stars in the \cnc\ and that the present star formation epoch is thus different from the epoch that formed the present $\sim70$ high-mass stars in Carina \citep{preibisch2011_laboca}. More optical jets, which come from sources close to the cloud surfaces, than IR jets from more deeply embedded sources, are seen in the region, showing that presently forming stars predominantly lie near the surfaces of clouds.

\begin{acknowledgements}
The authors would like to thank the \spitzer\ Science Center Helpdesk for their support with \mopex.

% DFG/Excellence Cluster
This work was supported by the German \emph{Deut\-sche For\-schungs\-ge\-mein\-schaft, DFG\/} project number 569/9-1. Additional support came from funds from the Munich Cluster of Excellence: “Origin and Structure of the Universe”.

% Spitzer
This work is based in part on archival data obtained with the \spitzer\ Space Telescope, which is operated by the Jet Propulsion Laboratory, California Institute of Technology under a contract with NASA.

% Herschel
This publication made use of data obtained with the \herschel\ spacecraft. \herschel\ is an ESA space observatory with science instruments provided by European-led Principal Investigator consortia and with important participation from NASA.

% 2MASS
This publication made use of data products from the Two Micron All Sky Survey, which is a joint project of the University of Massachusetts and the Infrared Processing and Analysis Center/California Institute of Technology, funded by the National Aeronautics and Space Administration and the National Science Foundation.

% DSS
This publication made use of data products from the Digitized Sky Survey, which was produced at the Space Telescope Science Institute under U.\,S.\ Government grant NAG W-2166. The images of these surveys are based on photographic data obtained using the Oschin Schmidt Telescope on Palomar Mountain and the UK Schmidt Telescope. The plates were processed into the present compressed digital form with the permission of these institutions. 

% ADS
This research made use of NASA's Astrophysics Data System Bibliographic Services.

\end{acknowledgements}

\bibliographystyle{aa}
%\bibliography{jetquellen.bib}

\end{document}